\documentclass[iicol,sn-mathphys-num]{sn-jnl}

\usepackage{graphicx}%
\usepackage{multirow}%
\usepackage{amsmath,amssymb,amsfonts}%
\usepackage{amsthm}%
\usepackage{mathrsfs}%
\usepackage[normalem]{ulem}
\usepackage[title]{appendix}%
\usepackage{xcolor}%
\usepackage{textcomp}%
\usepackage{manyfoot}%
\usepackage{booktabs}%
\usepackage{algorithm}%
\usepackage{algorithmicx}%
\usepackage{algpseudocode}%
\usepackage{listings}%

\theoremstyle{thmstyleone}%
\theoremstyle{thmstyletwo}%

\theoremstyle{thmstylethree}%

\begin{document}

\title[Classifying Complex Dynamical and Stochastic Systems via Physics-Informed Recurrence Features]{Classifying Complex Dynamical and Stochastic Systems via Physics-Based Recurrence Features}


\author[1,2]{\fnm{J. V. M.} \sur{Silveira}}\email{jorge.malosti@gmail.com}

\author[1,2]{\fnm{H. C.} \sur{Costa}}

\author[1,2]{\fnm{G. S.} \sur{Spezzatto}}

\author*[1,2]{\fnm{T. L.} \sur{Prado}}\email{thiagolprado@ufpr.br}

\author[1, 2]{\fnm{S. R.} \sur{Lopes}}

\affil[1]{\orgdiv{Department of Physics}, 
           \orgname{Federal University of Paraná}, 
           \orgaddress{\city{Curitiba}, \postcode{81531-980}, \country{Brazil}}}

\affil[2]{\orgdiv{Interdisciplinary Center for Science, Technology and Innovation (CICTI)}, 
          \orgname{Federal University of Paraná}, 
          \orgaddress{\city{Curitiba}, \country{Brazil}}}

\abstract{In this study, we employ the recently developed recurrence microstate probabilities as features to improve accuracy of several well-established machine learning (ML) algorithms. These algorithms are applied to classify discrete and continuous dynamical systems, as well as colored noise. We demonstrate that the dynamical characteristics quantified by this method are effectively captured in recurrence microstate space, a space defined solely by the recurrence properties of the signal. This space change reduces dimensions, which also reduces the time needed to perform calculations and obtain relevant information about the underlying system. Here, we also demonstrate that a few optimal machine learning (ML) algorithms are particularly effective for classification when combined with recurrence microstates. Furthermore, these new machine learning vectors significantly reduce memory usage and computational complexity, outperforming the direct analysis of raw data.}
\keywords{Recurrence microstates, Machine learning, Dynamical systems, Classification algorithms}

\maketitle

\section{Introduction}
\label{introduction}

Data mining, or the practice of analyzing datasets in order to generate new information, is crucial in today's world due to its ability to uncover hidden patterns such as system dependent parameters \cite{han2012data}. This practice enables the identification of trends, correlations, and anomalies that might not be immediately obvious, helping to optimize processes or increase efficiency. In technological situations, data mining helps to uncover actionable insights, improve decision-making, and drive innovation by transforming raw data into valuable knowledge. Modern data mining  techniques, such as machine learning, achieve optimal performance when the maximum amount of useful information is condensed into the smallest feasible phase space. Redundant information can impede these analyses, underscoring the importance of dimensionality reduction and meticulous data selection for achieving effective results \cite{maosaobra, haykin2009neural}. Consequently, a critical balance must be struck between data volume and machine learning capabilities, maximizing useful information while minimizing redundancy.

However, most physical systems possess many degrees of freedom, meaning they are represented in high-dimensional phase spaces. This high dimensionality often implies large volumes of data, which may contain significant redundancy among components. Nevertheless, it is rare for any individual coordinate in phase space to be entirely devoid of new information; each typically contributes unique insights and should therefore be included in some form  \cite{pathria2016statistical,STROGATZ2001}.

Dimensionality reduction algorithms are widely employed to improve the efficiency of data visualization, clustering, and classification tasks, both with and without machine learning \cite{velliangiri2019review}. By mapping datasets to a lower-dimensional space, these algorithms extract key features and enhance computational efficiency, as models perform better with smaller datasets. Techniques such as Principal Component Analysis (PCA) are utilized across various fields \cite{jolliffe2016principal}, generating orthogonal components that maximize data variance and improve interpretability while minimizing information loss. More advanced methods, such as Uniform Manifold Approximation and Projection (UMAP) \cite{mcinnes2018umap}, leverage Riemannian geometry and algebraic topology to reduce dimensionality while preserving global structure. This makes UMAP particularly effective for visualizing high-dimensional data in two or three dimensions, as well as for preprocessing inputs for machine learning algorithms.

In this context, it is crucial to consider space transformations that condense the information contained in the original data. Such transformations to a more physically oriented space where distinctness may be boosted simplify the process of extracting information through machine learning, enhancing its efficiency \cite{chandrashekar2014survey, MLDataTransformations}. However, it is essential that these transformations preserve as many characteristics of the original system as possible. In summary, modern machine learning aims for a compact yet informative phase space, where carefully selected features capture the essential dynamics without redundancy. Achieving this balance enhances computational efficiency, interpretability, and model performance \cite{jia2022feature}.

A particularly relevant space transformation is the one that leads to the recurrence space. Such a transformation does not require, for example, the stationarity condition of the data \cite{Marwan2007}, as is the case with Fourier transforms. Recurrence analyses can also be applied to periodic or non-periodic, linear or nonlinear systems. The fundamental principle of this technique is anchored in Poincaré's recurrence theorem \cite{Poincare1890}, which states that a given dynamical system (excluding extreme solutions that diverge to infinity or converge to a fixed point) must eventually return arbitrarily close to a past state $i$, provided that sufficient time is given. Poincaré's fundamental idea was later used to develop trajectory analysis methods, among which recurrence plots stand out \cite{Eckmann1987}. This tool transforms the properties of a trajectory into a binary matrix of possible states, where a digit $1$ in element $i,j$ of this matrix indicates that states $i$ and $j$ of the trajectory recur, while a digit $0$ means that there is no recurrence between the considered states. Many measures have been developed to quantify the various patterns that emerge in this matrix \cite{zibilut2007}; the complete set of these measures constitutes the analysis of recurrence quantifiers \cite{Marwan2007, Marwan2023}.

The recurrence matrix of individual states in a trajectory can be generalized into the concept of recurrence microstates \cite{Corso2018, Prado2020}, which describe the recurrence patterns of sequences of $N$ values in a trajectory. In this framework, the binary digit representing two possible recurrence states yes (1) or no (0) -- as described by Eckmann \cite{Eckmann1987} for the recurrence between two individual states, is extended to $2^{N^2}$ possible recurrence patterns of sequences. In other words, it becomes a binary sequence of $N^2$ digits \cite{Corso2018, Prado2020}.

Here, we use the transformation of real data into recurrence microstate distributions as a way to encode the information of a physical system. These microstate methodologies have already been applied in various contexts, such as the analysis of dimensional transitions in magnetic materials \cite{corso2021maximum}, electrocardiogram data \cite{boaretto2024use}, and human electroencephalogram data \cite{ferre2024cycling}, as well as in sound recording analysis \cite{Prado2021}. In a different approach, the same microstate probabilities can be used to construct a Microstates Multi-Layer Perceptron (MMLP) method as a supervised classifier for time series from chaotic dynamical systems \cite{spezzatto2024recurrence}. It is observed that the use of the space defined by the probabilities of occurrence of each possible recurrence microstate as pre-processed data improves the efficiency of machine learning classification methods \cite{spezzatto2024recurrence}. We consider ten well-known machine learning methods \cite{maosaobra, scikit-learn, introdatasci}: Decision Tree, Random Forests, K-Nearest Neighbors (KNN), Support Vector Classifier (SVC), Linear SVC, Gaussian Naive Bayes (Gaussian NB), Bernoulli Naive Bayes (Bernoulli NB), Gradient Boosting, Multi-Layer Perceptron (MLP), and Logistic Regression. All these methods are tested on seven chaotic dynamical systems, namely five different discrete datasets: the $\beta x ,\mathrm{mod(1)}$ (a generalization of the Bernoulli shift), logistic, Gauss, Hénon, and Ikeda maps, as well as two examples of continuous coupled differential equation systems, the Lorenz and Rössler oscillators \cite{STROGATZ2001, ChaosandNonlinear}. The methods are also applied to different types of time-correlated stochastic noise (colored noise) \cite{Weissman1988}. In all situations, our fundamental question is: Correctly classify the parameter used in the generation of the time series.

In Section 2, we present the essential concepts of recurrence matrices and recurrence microstates, along with the notion of maximum entropy. In Section 3, we discuss the different machine learning methods used, as well as the definitions of hyperparameters. In Section 4, we describe the methodology applied in our analyses. In Section 5, we present the results obtained when using recurrence microstates as pre-processed data. In Section 6, we show the results obtained from direct data analysis, without using recurrence microstates. Finally, in Section 7, we provide a discussion and conclusion.

\subsection{Recurrence Microstates}

Recurrence is a fundamental property of ergodic and stationary systems, playing a critical role in the analysis of chaotic and stochastic systems \cite{STROGATZ2001,Marwan2007}. The key concept behind recurrence is the Poincaré recurrence theorem, which states that in ergodic dynamical systems, after a sufficiently long period, the system will eventually return to a state that is arbitrarily close to its initial condition \cite{Poincare1890}.

Eckmann {\textit{et al.}} \cite{Eckmann1987} proposed a tool to visualize recurrence properties of a length $K$ trajectory, the so called recurrence plots (RP). These plots consist of  binary $K\times K$ matrices, that compare different individual points in time of a state vector in phase space, identifying recurrent (nonrecurrent) states as ``1" (``0")  states. By encoding a recurrent state as  black pixels while nonrecurrent as white pixels, the recurrence plot draws a mosaic of black and white patterns, which translates the recurrence characteristics of a trajectory. Each element of the recurrence matrix is then defined by 
\begin{equation}
\label{equation:rp}
\text{R}_{ij}(\epsilon) = \Theta\left(\varepsilon - \|\vec{x}_i - \vec{x}_j\|\right), \quad i, j = 1, \dots, K,
\end{equation}
where $K$ is the size of the trajectory $\vec{x}$, \, $\varepsilon$ is the recurrence threshold, $\Theta(\cdot)$ is the Heaviside function (that is, \(\Theta(x) = 0\) if \(x < 0\), and $\Theta(x) = 1$ otherwise) and \(\|\cdot\|\) represents a suitable  metric to be used to calculate the distance among states \cite{Marwan2007}. 

\begin{figure}[hbtp]
\centering
\includegraphics[width=1.0\linewidth]{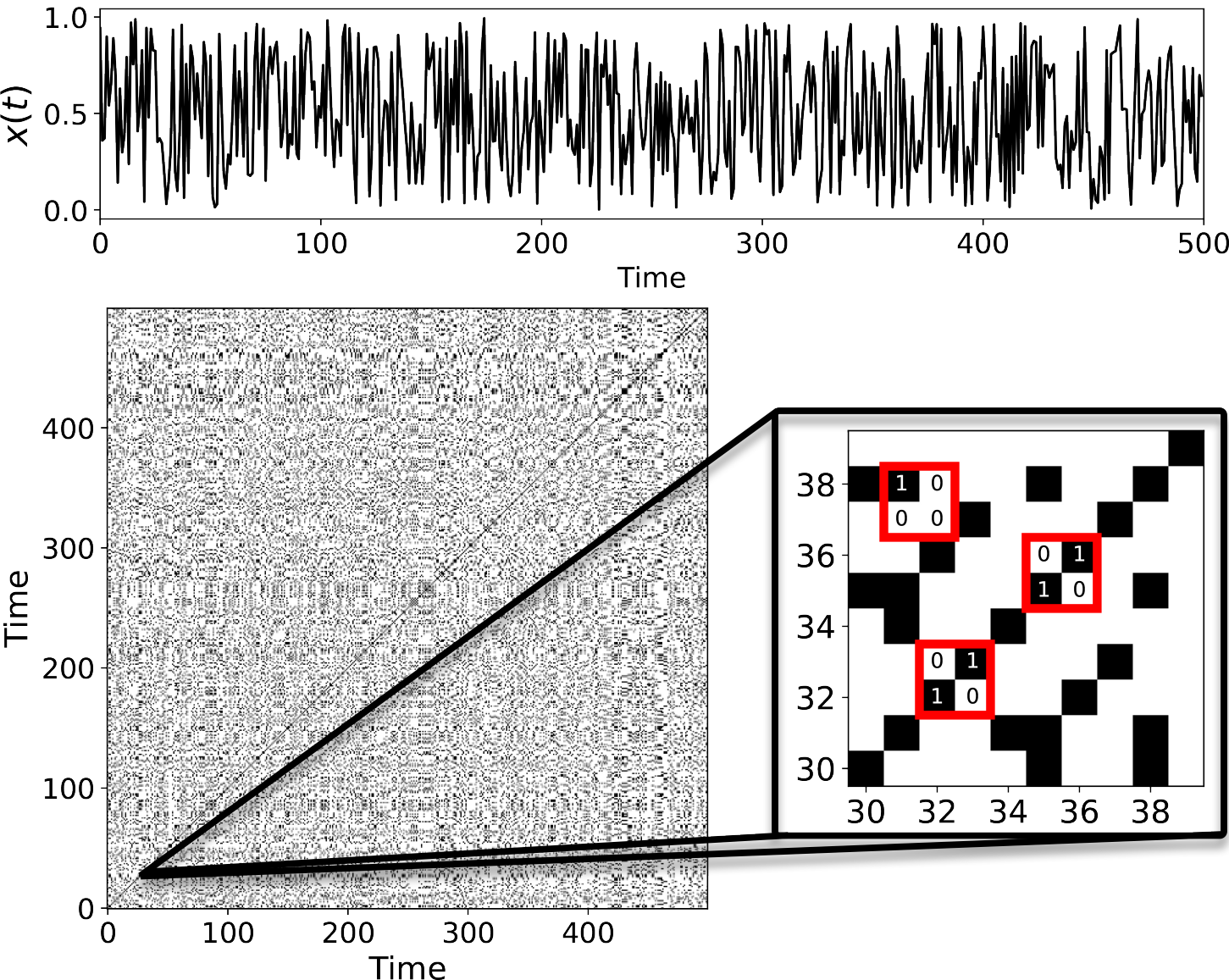}
\caption{Example of microstates embedded in a $(K\times K)$ RP. A short data sequence of size $N < K$ is translated into a recurrence microstate, a $(N\times N)$ matrix encoding recurrence relations of short sequences of the data. In the figure, the $2\times2$ microstates are highlighted in red. Three such microstates can be observed, with those along the diagonal corresponding to the same configuration.}
\label{fig:esquema-microestados}
\end{figure}

Here we focus on the concept of Recurrence Microstates, as proposed by Corso \textit{et al.} \cite{Corso2018}, which involves extracting an ensemble of $N \times N $ submatrices from the recurrence matrix. Since the RP is a binary matrix composed of ones and zeros, the microstates are formed by all possible combinations of these binary submatrices. These microstates capture different recurrence configurations present in the RP, providing a rich way to characterize the underlying dynamics of the analyzed time series. To make recurrence microstates clear,  Fig. \ref{fig:esquema-microestados} shows three examples of them embedded in the entire RP for the simplest $N=2$ case, namely $2\times 2$ matrices.

Given a $K \times K$ recurrence matrix, we can extract up to $M$ distinct $N \times N$, $N<K$, recurrence microstates. The maximum possible value of $M$ is $(K-N+1) \times (K-N+1)$, which accounts for the overlapping microstates formed when sliding an $N \times N$ window across the entire recurrence matrix.

The recurrence matrix depends on the recurrence threshold $\varepsilon$, as shown in Eq. \ref{equation:rp}. Following the approach in \cite{Prado2020}, we determine the optimal value of $\varepsilon$ by maximizing the entropy of recurrence microstates. This entropy is computed as a function of the probabilities of occurrence for each microstate and the recurrence threshold $\varepsilon$.
\begin{equation}
    S(\varepsilon)=-\sum_{i=1}^{2^{N^{2}}}P_{i}(\varepsilon)\ln \left[P_{i}(\varepsilon)\right],
    \label{entropia}
\end{equation}
such that $\mathrm{max\,}[ S(\varepsilon)]$ turns out to be $S_{max},$ uniquely defining the value of $\varepsilon$ all along this article.

Therefore, in this work, the recurrence threshold $\varepsilon$ is not treated as an arbitrary tuning parameter but as the one that yields the most informative and diverse set of recurrence microstates. By selecting $\varepsilon$ that maximizes the entropy $S(\varepsilon)$, we ensure that the resulting microstate distribution captures the largest possible variety of dynamical patterns present in the data, thus representing the system’s complexity in an optimal way.

\subsection{Machine Learning Methods}

Artificial intelligence (AI), along with one of its most significant branches, machine learning (ML), has become a game-changer across many fields of science and technology. Since the early work on the perceptron in the 1940s and 1950s \cite{mcculloch1943logical}, often considered one of the foundational models of neural networks, the field has experienced tremendous growth, evolving from theoretical concepts to practical, real-world applications in numerous domains. This growth has been driven by several key factors, including the exponential increase in computational power, the availability of large datasets, and advances in algorithms. 

\textit{In this study, we evaluate the performance of several well-known classification methods using recurrence microstates as the sole features}. The probability distributions of recurrence microstates, derived from the data, are used to preprocess the information before being fed into each machine learning (ML) method. This approach allows us to assess the effectiveness of each method. Furthermore, we investigate which methods achieve the highest accuracy while maintaining low computational costs in terms of processing power. Brief descriptions of the ML methods used in this analysis are provided in Table \ref{tab1}. The machine learning methods used in this work are from the python package scikit-learn  \cite{scikit-learn} and keras \cite{chollet2015keras}.
\begin{table*}[!hbtp]
    \small
    \centering
    \begin{tabular}{| c | p{11cm} |}
    \hline 
    \textbf{Methods} & \makebox[11cm][c]{\textbf{Principles}} \\
    \hline\hline
    \multirow[c]{2}{*}{\textbf{Decision tree}} & It uses data samples to create decision rules in the form of a tree, making it easier to understand and apply these rules in decision-making. \\
  
    \hline
    \multirow[c]{2}{*}{\textbf{Random forests}} & A set of Decision Trees trained on random subsets of the same dataset. For predictions, each tree provides its own estimate, and the final class is determined by the majority vote among all the trees.\\
   
    \hline
    \multirow[c]{2}{*}{\textbf{KNN}} & It identifies the nearest training samples to a new point based on distance and predicts the label using these samples, assuming that nearby examples are similar.\\

    \hline
    \multirow[c]{2}{*}{\textbf{SVC}} & Classifies a test observation based on which side of a hyperplane it falls on, with the hyperplane chosen to optimally separate the majority of training observations into two classes.\\

    \hline
    \multirow[c]{2}{*}{\textbf{Linear SVC}} & Using a linear kernel, it identifies a separating hyperplane that maximizes the margin, which is the minimum distance between data points of different classes and the decision hyperplane.\\

    \hline
    \multirow[c]{2}{*}{\textbf{Gaussian NB}} & It uses Bayes' Theorem with conditional probabilities, assuming attribute independence and that continuous variables follow a Gaussian distribution.\\

    \hline
    \multirow[c]{3}{*}{\textbf{Bernoulli NB}} & It calculates the probability of an example belonging to each class based on the presence or absence of features in multivariate Bernoulli distributions, multiplying these probabilities to select the class with the highest probability as the prediction.\\

    \hline
    \multirow[c]{2}{*}{\textbf{Gradient Boosting}} & It adds predictors sequentially to a set, with each new predictor correcting the errors of its predecessor, aiming to adjust the new predictor to fix the residual errors made by the previous one.\\

    \hline
    \multirow[c]{3}{*}{\textbf{MLP}} & It consists of multiple layers of artificial neurons arranged hierarchically, including an input layer, one or more hidden layers, and an output layer. Neurons in each layer are connected to the next by adjustable weights, and each connection has a non-linear activation function.\\

    \hline
    \multirow[c]{2}{*}{\textbf{Logistic Regression}} & It calculates the probability of a binary dependent variable (with two possible outcomes) based on one or more independent variables using the logistic function.\\
    
    \hline
    \end{tabular}
\caption{Main characteristics of machine learning methods used in the work. \cite{maosaobra}\cite{scikit-learn}\cite{introdatasci}}
\label{tab1}
\end{table*}

Our objective is to present a clear and effective approach for achieving optimal results by leveraging recurrence microstate data to power well-established AI classification algorithms. For this reason, we focus exclusively on testing these classical algorithms. While other methods may potentially achieve higher success rates, they are often less accessible and will not be evaluated in this study.

\section{Methodology}

We evaluate the performance of various machine learning algorithms in classifying parameters of chaotic dynamical systems and the degree of temporal correlation in stochastic noise. Specifically, we examine five discrete-time chaotic systems: the $\beta x \mod(1)$, logistic, Gauss, Hénon, and Ikeda maps, along with two continuous-time systems: the Lorenz and Rössler oscillators. In addition, we investigate several types of colored noise, which, despite being stochastic processes, exhibit distinct temporal correlations. For each system---whether maps, flows, or colored noise---we generate a set of 40 time series, each corresponding to a different parameter selected from pre-defined intervals.

\subsection{Time series generation}

For each discrete (continuous) dynamical system, $40$ time series of $1,000$ ($3,000$) data points are generated for each of the $40$ pre-selected parameters, which are uniformly distributed within pre-defined intervals as may be observed in Table \ref{tab:params} which also brings details of the parameter ranges for each system. In total, we generate $1,600$ time series for each system. To ensure that the time series accurately represent the stationary chaotic behavior of the systems, each series is generated after discarding the first $1,000$ transient data points.
\begin{table*}[!htpb]
\centering\small
\begin{tabular}{|c|c|c|c|} 
\hline
\textbf{System} & \textbf{Type}    & \textbf{Parameter Range}       & \textbf{Fixed Parameters}      \\ \hline\hline
$\beta x$       & Map              & $1.99 \leq \beta \leq 6.99$    & $-$                            \\ \hline
Logistic        & Map              & $3.60 \leq r \leq 4.00$        & $-$                            \\ \hline
Gauss           & Map              & $-0.70 \leq \gamma \leq -0.30$ & $\phi = 6.20$                \\ \hline
Henon           & Map              & $1.10 \leq a \leq 1.20$        & $b = 0.30$                     \\ \hline
Ikeda           & Map              & $0.60 \leq u \leq 0.89$     & $-$                            \\ \hline
Lorenz          & Flow             & $27.99 \leq \rho \leq 37.99$   & $\sigma = 10$, $\beta = 8/3$   \\ \hline
Rossler         & Flow             & $0.20 \leq a \leq 0.30$        & $b = 0.2$, $c = 5.7$           \\ \hline
\end{tabular}
\caption{Range of parameters for each discrete and continuous dynamical system.}
\label{tab:params}
\end{table*}

In addition to the previously discussed discrete and continuous deterministic systems, all machine learning methods are also tested to evaluate properties of time correlation in  stochastic noises. This type of noise is characterized by a non-zero autocorrelation function, which indicates statistical dependence between noise values at different time scales.

The non-zero autocorrelation function observed in colored noise is reflected in its frequency domain representation, resulting in a power law relationship.  The power spectral density follows a $1/f^\alpha$ distribution, with $\alpha$ varying (here) from $-2$ to $2$, where $\alpha = 0$ corresponds to white noise. So, for colored noise, the power spectral density decreases as the frequency increases. To generate the stochastic data, we use the methodology proposed by Thieler \textit{et al.} \cite{thieler2016robper}, adapting the implementation from  \texttt{R} \cite{R2024} to \texttt{Julia} \cite{Julia17} language.

\subsection{Recurrence microstates entropy and recurrence microstates}

From the time series data for each scenario, a recurrence analysis is performed. Specifically, a sufficiently large set of recurrence microstates is randomly sampled from the original data, resulting in a set of microstate probabilities. If $N=2$, this set comprises 16 microstate probabilities. For $N=3$, the set reflects 512 different microstate probabilities. In general, for any $N$, $2^{N^2}$ different probabilities are obtained. Additionally, this probability set is used to generate an integral quantifier of the data: the recurrence microstate entropy given in Eq. \ref{entropia}. This method enables a significant reduction in data size while potentially increasing the amount of useful information for distinguishing between datasets with different parameters.

\section{Results}
\subsection{Accuracies of ML algorithms using only recurrence microstates data}

\begin{figure*}[htb]
    \centering
    \includegraphics[width=\textwidth]{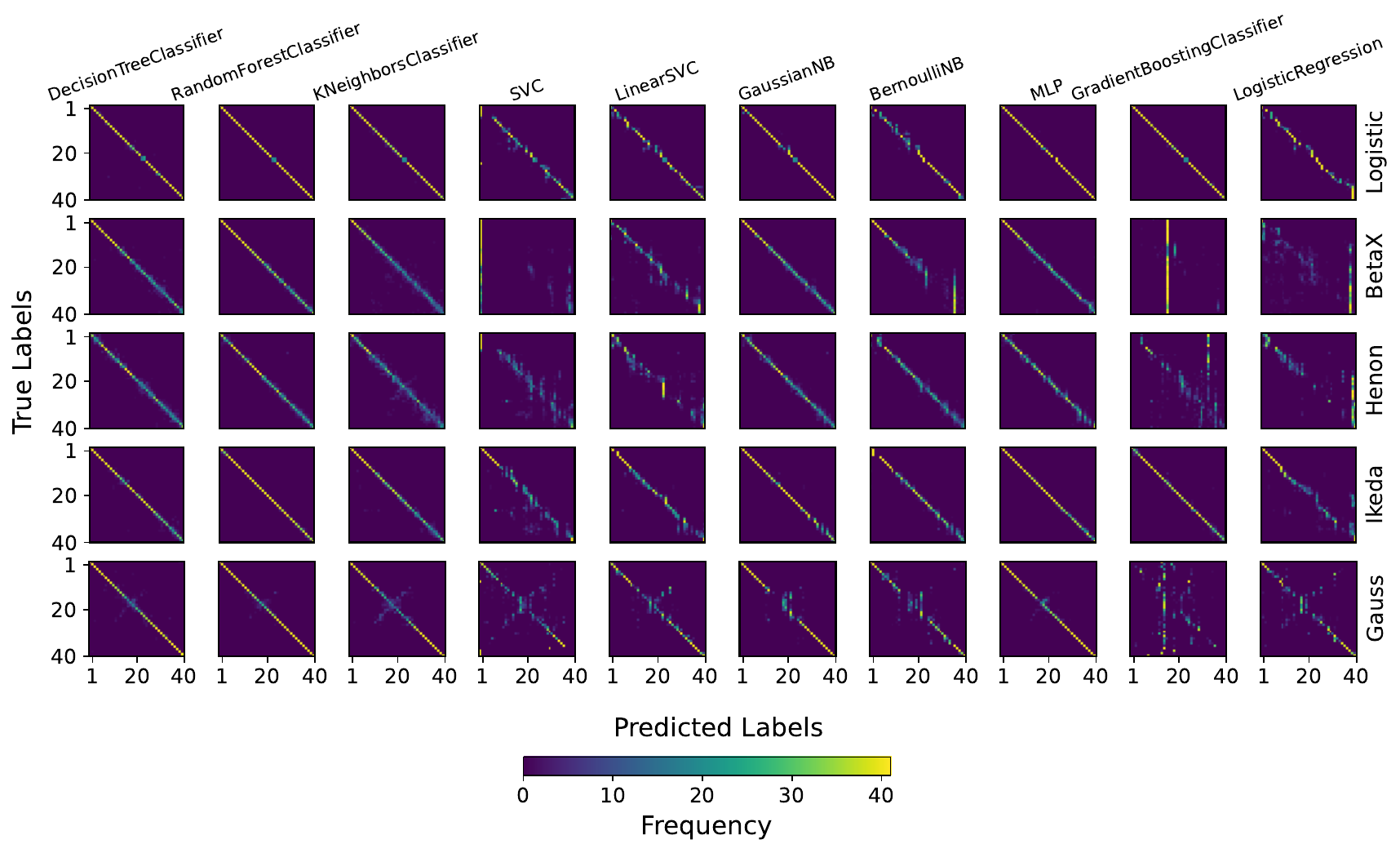} 
\caption{Confusion matrices for predictions by various machine learning models classifying parameters of selected dynamical maps. Each column corresponds to a specific classification model, while each row represents a dynamical map. The horizontal axis shows predicted labels, and the vertical axis shows true labels. The color scale indicates prediction frequencies, with yellow shades representing higher counts. The colorbar below provides the absolute counts for each cell in the matrices. Values along the main diagonal correspond to correct classifications, indicating the number of times each class was accurately predicted by the model.}

    \label{fig:FIG2}
\end{figure*}

As first results, we present the confusion matrices of all ML methods using only the sets of microstates obtained from time series of discrete dynamical systems (iteration maps) studied in this work. A confusion matrix summarizes the performance of a classification model by displaying the number of correct and incorrect predictions for each class, with rows representing the true classes and columns representing the predicted ones. Fig.~\ref{fig:FIG2} shows the confusion matrices for the machine learning methods used in the classification of the parameters of the maps for microstates of size $N=3$

In general, the set of microstates obtained for each time series (512 features) adequately translates the behavior of all maps for each one of the $40$ different parameters. This fact is reflected in the clear diagonals obtained for most of the learning algorithms. Almost in all panels, the errors made by the algorithms result in small differences between the true and predicted values, leading to the appearance of secondary (short) diagonals alongside the main diagonal in most panels of Fig~\ref{fig:FIG2}. However, some methods completely fail to achieve adequate predicted results, such as SVC applied to the $\beta x$ and Gauss maps, as well as the Gradient Boosting Classifier when applied to the $\beta x$, Henon, and Gauss maps. The Logistic Regression algorithm also does not yield good results when applied to the representative microstate sets of the $\beta x$ map. It is also interesting to note that some algorithms still exhibit failures in identifying parameter intervals of the maps, as can be observed with the Bernouilli Naive Bayes algorithm when applied to the dynamics for some ranges of $\beta$ in the $\beta x$ map.

Since other algorithms succeed in all cases, our conclusion is that the information contained in the sets of microstates is preserved and it is consistent with the time series across all parameter intervals. The failure of a particular algorithm is due to its own evaluation methods and not to the absence of information.

A visual inspection of Fig. 2 shows that the Random Forest Classifier algorithm achieves the best results. At first glance, this fact suggests choosing this method as the ideal one. However, other variables may also be important, such as the computational time required for each of the algorithms. We will discuss these details further ahead.

The Support Vector Classifier  (SVC) and Bernoulli Naive Bayes methods showed unsatisfactory performance in this work due to specific limitations of each model \cite{haykin2009neural, wickramasinghe2021naive}. SVC, for example, has high computational complexity, which makes it difficult to scale for large datasets with high dimensionality. Additionally, the choice of kernel is a decisive factor for performance. Although the RBF (Radial Basis Function) kernel was used, it proved to be inefficient for the data in question, possibly due to the high dimensionality of the features and the difficulty in properly tuning the hyperparameters.  On the other hand, the LinearSVC model, which uses a linear kernel, performed better. This can be attributed to the fact that in high-dimensional spaces, classes tend to become approximately linearly separable \cite{cover1965geometrical, haykin2009neural}, which benefits linear algorithms. Logistic Regression, although also linear, had slightly inferior performance compared to LinearSVC. This behavior can be explained by the fact that Logistic Regression maximizes conditional probability, while LinearSVC optimizes the maximum margin between classes, making it more robust in scenarios with noise or close boundaries. Nevertheless, both outperformed the non-linear SVC when the number of features increased, for microstates of size $N=3$ and $N=4$, highlighting the efficiency of linear methods in high-dimensional spaces. As for Bernoulli Naive Bayes, which requires samples to be represented as binary vectors, it is more suitable for data with binary characteristics, such as in text classification tasks. Although the algorithm automatically converts features to binary values based on thresholds, this simplification is not suitable for the data in this study, which have complex continuous distributions. However, when using microstates of size $N=3$ and $N=4$, Bernoulli Naive Bayes achieved more significant accuracies. This can be attributed to the fact that microstates of larger sizes have more distinct distributions, making the separation between classes more evident even under the binary approach.

Gradient Boosting is an algorithm that is highly sensitive to various factors, such as the choice of training parameters, the size of the data, and the specific characteristics of the problem. This sensitivity may explain its performance variability. In some cases, the model adapts well to the data and is able to capture the complex relationships between the features, resulting in excellent performance. However, when conditions change, such as an increase in the number of features, the model may exhibit inferior or even unstable performance. Additionally, Gradient Boosting is prone to overfitting if the parameters are not carefully tuned, which also contributes to inconsistent results across different runs. This variability, therefore, reflects the complexity of the algorithm, which requires careful and specific configuration to optimize its performance.
\begin{figure}[ht]
    \centering
    \includegraphics[width=1.0\linewidth]{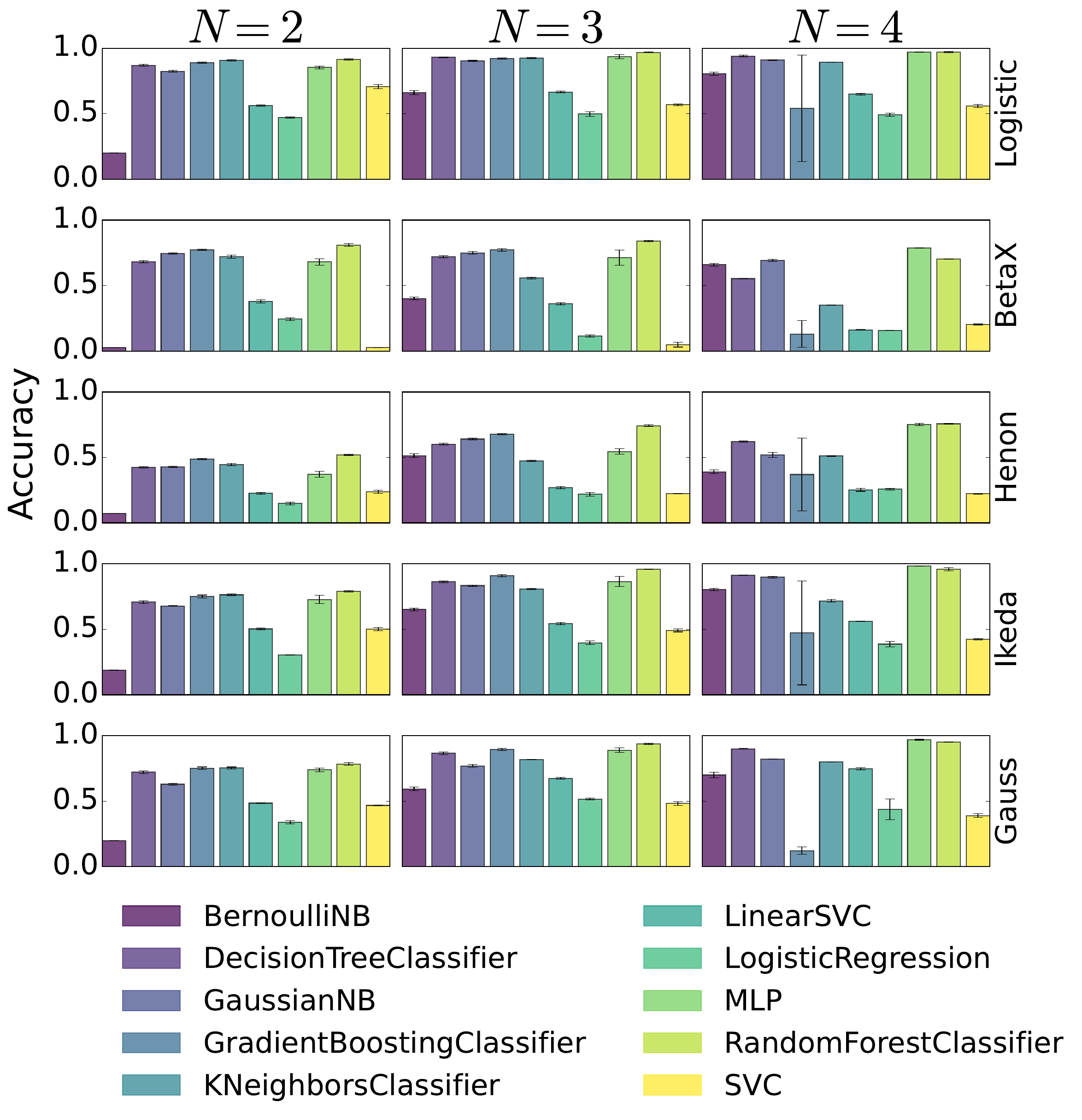}
    \caption{Mean Accuracy achieved by various machine learning algorithms for predicting each dynamic system. Results are displayed for N=2, 3 and 4, where $N$ represents the size of the recurrence microstates. Each bar represents the mean accuracy obtained by running all 20 ordered train–test permutations across five distinct generated data sets. Error bars indicate the standard deviation.}
    \label{fig:acc_mapas}
\end{figure}

The panels displayed in Fig.~2 provide a visual representation of the performance of each classifier, indicating its success or failure in predicting the system parameters. In general, most methods, even when they do not correctly classify the parameters, exhibit only minor deviations from the true values. This behavior is visible in the diagonal-like patterns parallel to the main diagonal, showing that the predicted parameters, although not exact, remain close to the correct ones. To quantify the performance, Fig.~\ref{fig:acc_mapas} presents the classification accuracy, defined as the ratio between the number of correct predictions and the total number of predictions, for each machine learning method as a function of microstate size $N$. It is evident that increasing $N$ makes the microstate sets more representative of the underlying dynamics, leading to higher accuracy across nearly all maps. Even the methods that perform less effectively show a noticeable improvement for larger $N$.

It is expected that the size of the microstates plays a significant role in characterizing the dynamic features of the time series, as for $N=2$, these features are compressed into only $16$ microstates. For $N=3$, this set grows to $512$, allowing for a greater diversity of features to be captured. Such an increase in the size of the microstates leads to a higher computational cost. However, as evidenced in Fig. \ref{fig:acc_mapas}, even an extreme compression of the time series into a set of only $16$ quantifiers $(N=2)$  yields satisfactory results for most of our examples. This fact highlights the ability of the microstate set, even the smallest one, to effectively compress the characteristics of the time series.

\begin{figure}[ht]
    \centering
    \includegraphics[width=1.0\linewidth]{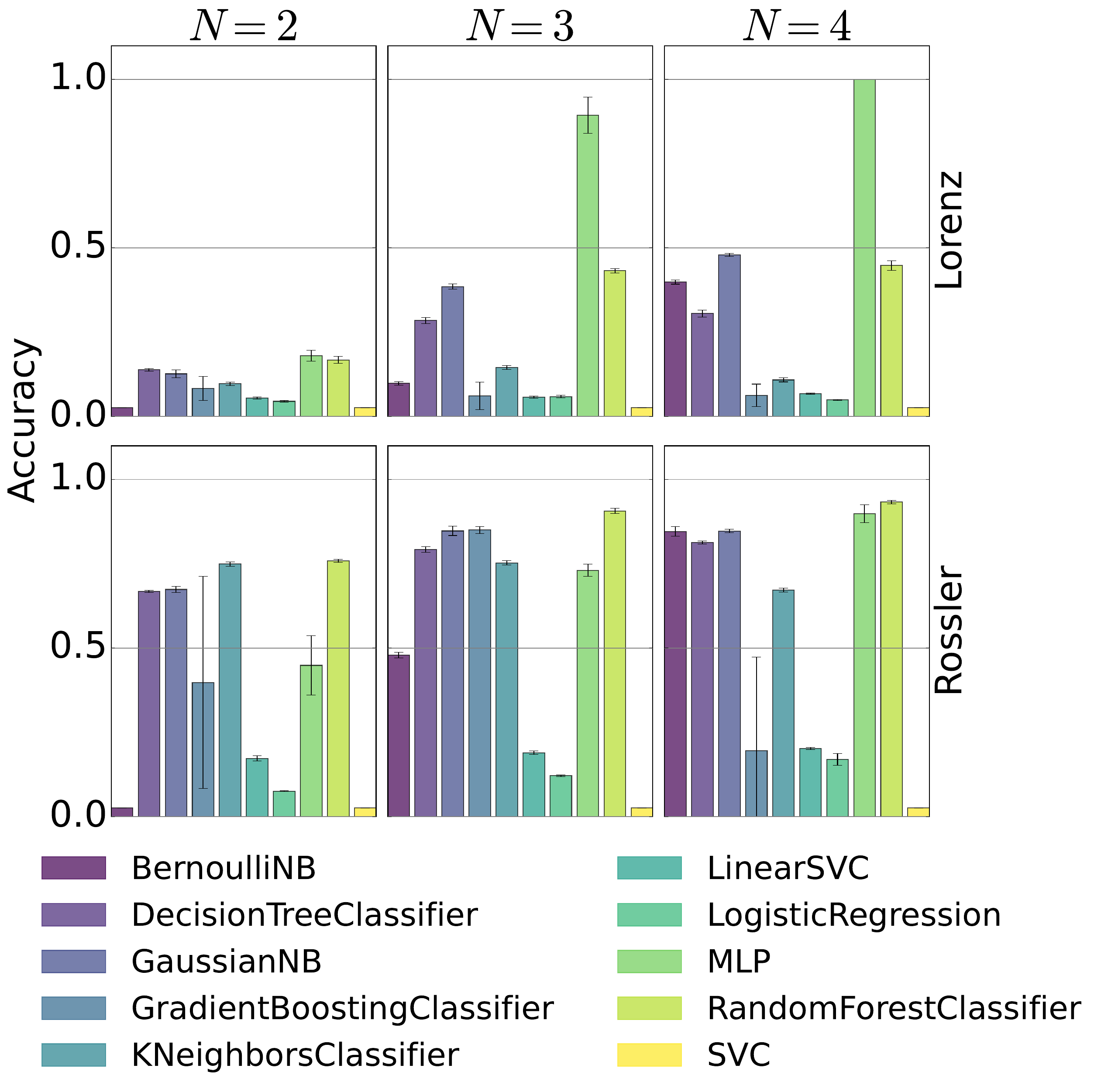}
    \caption{Machine learning accuracy for classifying two distinct continuous and chaotic dynamical systems.}
    \label{fig:acc_lorenz_rossler}
\end{figure}

The methods that demonstrated the best performance in this study are the MLP and RandomForest, both exhibiting consistent and satisfactory results in the classification of continuous and discrete systems, as well as in time correlated noise distinction. Additionally, an increase in accuracy is observed as the size of the microstates grows. Notably, for the Lorenz system, see Figure \ref{fig:acc_lorenz_rossler}, with $N=4$, the MLP achieved perfect performance, reaching 100\% accuracy even after averaging over five different training and test sets. On the other hand, Random Forest proves to be a more computationally efficient and easier-to-implement alternative compared to the MLP. The strong performance of these two models may be attributed to their capacity to handle complex, nonlinear interactions among features in the high-dimensional probability space defined by the recurrence microstates. After the entropy maximization procedure, the resulting probability distributions become highly informative and system-specific, thus, ensemble models such as Random Forest are particularly effective in mitigating fluctuations caused by finite data and recovering the underlying discriminative structure. In the case of MLP, despite its more opaque inner workings, its ability to approximate nonlinear manifolds may similarly capture the distinctive organization of microstate probabilities across different dynamical regimes. Intermediate methods, such as DecisionTree, GaussianNB, and KNN, achieved accuracies in the range of 60\% to 70\% for discrete systems, occasionally reaching performance levels comparable to the MLP in some cases. However, these methods exhibit greater variability in continuous systems, with less consistent performance in flows, making them less ideal for generalized applications.

Similarly to the approach taken for discrete dynamical systems, Fig. \ref{fig:acc_lorenz_rossler} presents the accuracies of the classifiers for two paradigmatic flows: the Lorenz and Rössler systems. In this context, an additional point is of importance. For an effective recurrence analysis, an appropriate temporal sampling of the flows is necessary. Sampling at very small time intervals leads to long time series and an excessive definition of the pseudo-periods of the series. On the other hand, sampling with too low resolutions may cause details of the dynamics to be lost. Here, we adopt the following criterion for selecting the sampling of continuous dynamical systems: (\textit{i}) we sample the systems as a function of the temporal resolution; (\textit{ii}) we compute the recurrence microstates entropy; and (\textit{iii}) we choose the temporal sampling that leads to the maximum entropy. The accuracies of the classification algorithms applied to the flows shown in Fig. \ref{equation:rp} demonstrate that the size of the microstate is an important variable in these cases. It is observed that for the Lorenz system, \(N=2\) proves inadequate for all tested algorithms. For \(N=3\) and \(N=4\), MLP emerges as the only suitable method. This behavior is not reflected in the analysis of the Rössler system, which shows several algorithms with satisfactory accuracies, although MLP still remains the ideal method for smaller \(N\).

\begin{figure}[ht]
    \centering
    \includegraphics[width=1.0\linewidth]{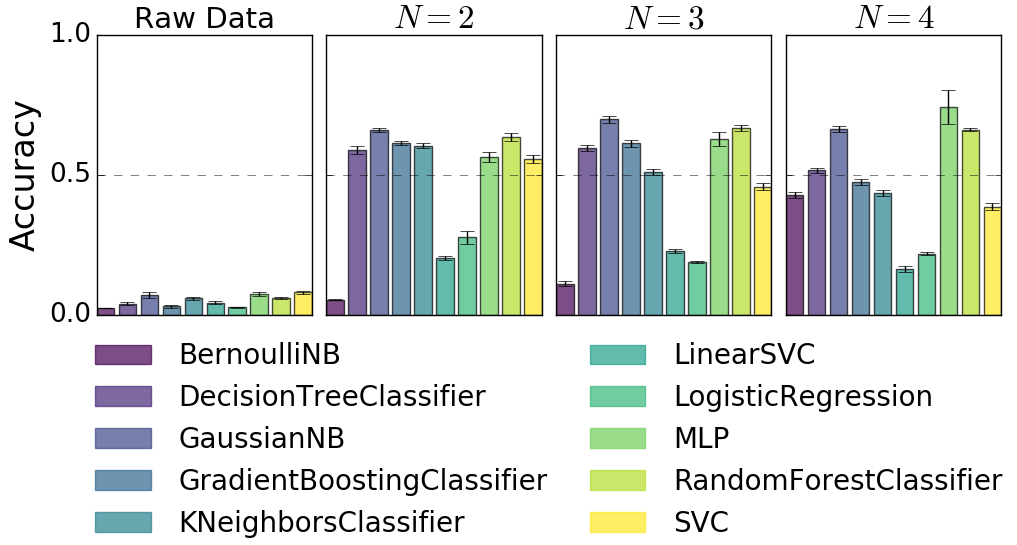}
    \caption{Machine learning accuracy for classifying color noises using the raw data and using the microstates analysis for $N=2$, $N=3$ and $N=4$. }    \label{fig:acc_noises}
\end{figure}

\subsection{Accuracies of ML algorithms using only the data}

\begin{figure}[htb]
    \centering
    \includegraphics[width=1.0\linewidth]{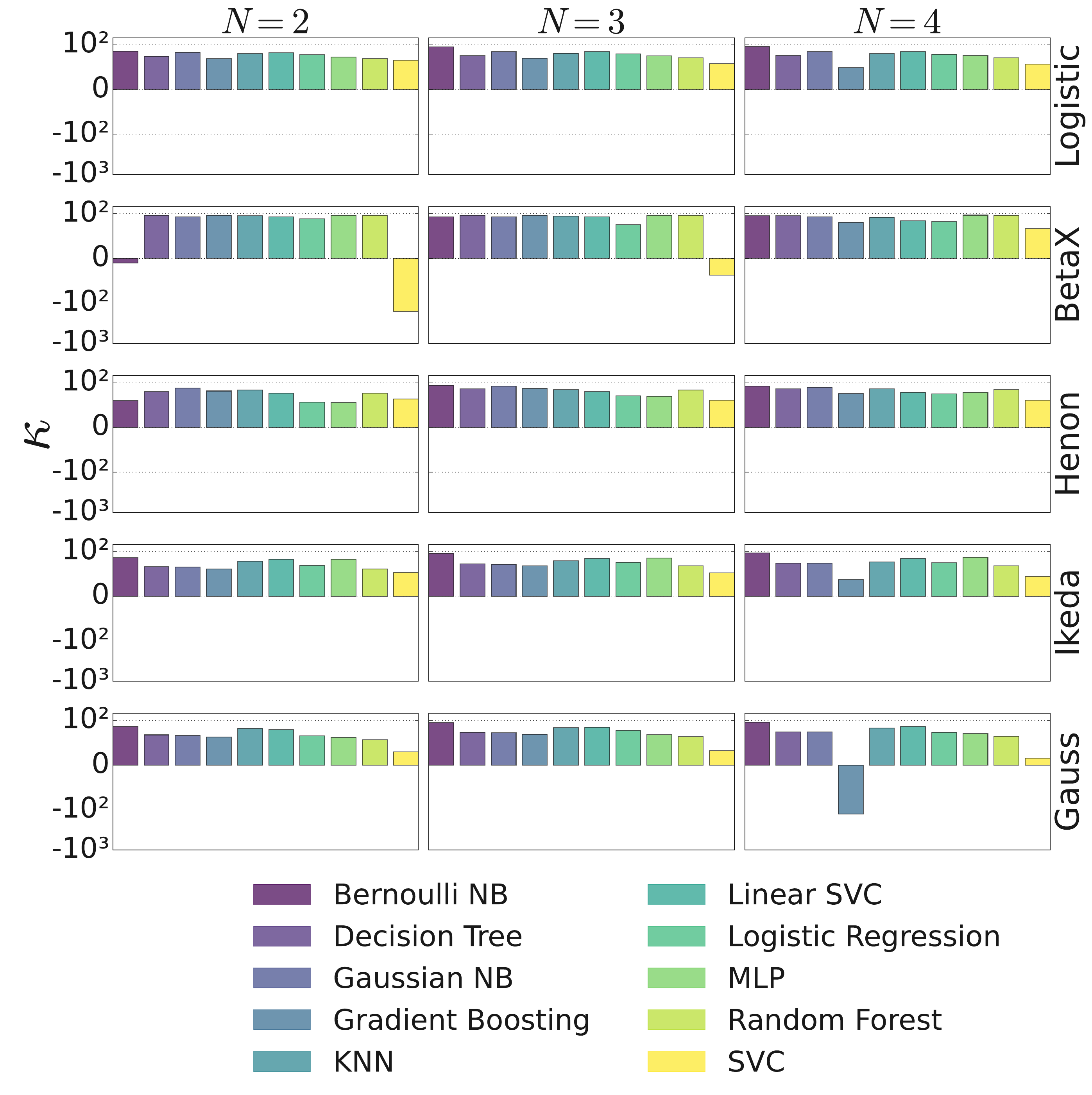}
    \caption{Measure of how better (or worse -- negative values) $(\kappa)$ calculates are the accuracies of pre-processed data using recurrence microstates  compared to the analyses done using raw data in the context of discrete-time chaotic systems. This is given by the formula: \(\kappa = [({A_{\text{rec}} - A_{\text{raw}}})/{A_{\text{rec}}}] \times 100\), where \(\ A_{\text{rec}}\) is the mean accuracy obtained using recurrence microstates and \(\ A_{\text{raw}}\) is the mean accuracy obtained using raw data. This formula allows for a direct comparison between the accuracy achieved with recurrence microstates and raw data, clearly indicating the effectiveness of using recurrence microstates in numerical experiments across five discrete-time chaotic systems.}
    \label{fig:razao_mapas}
\end{figure}

\begin{figure}[!h]
    \centering
    \includegraphics[width=\linewidth]{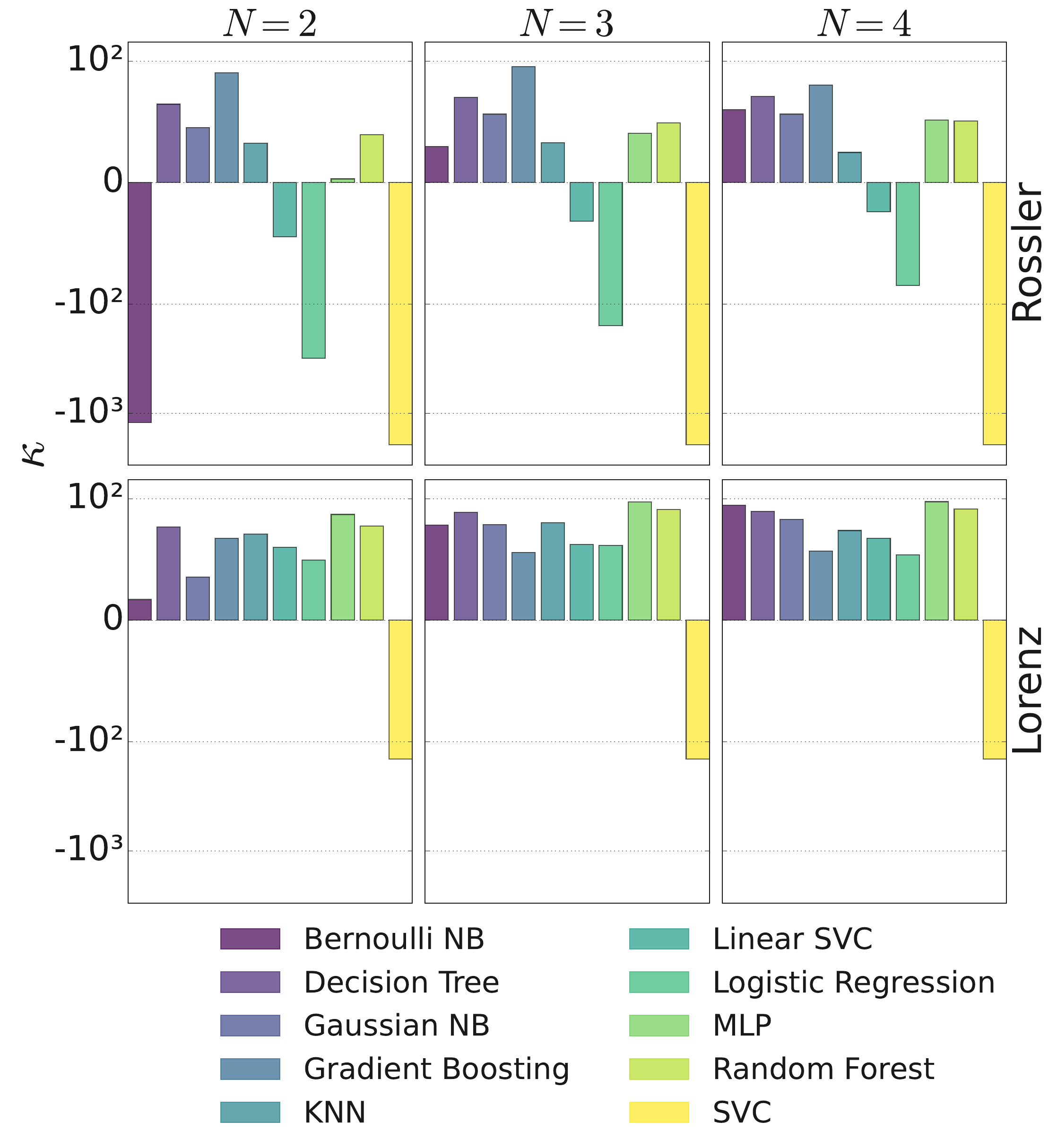}
    \caption{Measure of how better (or worse -- negative values) $(\kappa)$ calculates are the accuracies of pre-processed data using recurrence microstates  compared to the analyses done using raw data in the context of continuous-time systems. This is given by the formula: \(\kappa = [({A_{\text{rec}} - A_{\text{raw}}})/{A_{\text{rec}}}] \times 100\), where \(\ A_{\text{rec}}\) is the mean accuracy obtained using recurrence microstates and \(\ A_{\text{raw}}\) is the mean accuracy obtained using raw data. This formula allows for a direct comparison between the accuracy achieved with recurrence microstates and raw data, clearly indicating the effectiveness of using recurrence microstates in numerical experiments across five discrete-time chaotic systems.}
    \label{fig:razao_fluxos}
\end{figure}

Now we perform direct data analyses, considering machine learning methods classification directly on the raw data. In these cases we do not transform any data to the microstate-based space. In Fig. \ref{fig:acc_noises}, the first column displays the machine learning accuracies for the time-correlated noise raw data, demonstrating that these accuracies are, in general, substantially lower than those achieved when using microstate preprocessing. Figures \ref{fig:razao_mapas} and \ref{fig:razao_fluxos} compare the accuracies of machine learning methods for raw data and preprocessed data for maps and flows, respectively, for microstates of different sizes. As we can observe, overall, the classification results using raw data are significantly inferior to those obtained with preprocessing, regardless of the machine learning method used. However, there are some cases where the accuracy of the raw data is slightly higher, but in these cases, the overall accuracy of the method was already unsatisfactory, as occurred with SVC and BernoulliNB, as discussed earlier. Although the time required to perform recurrence microstate preprocessing should not be overlooked, it does not pose a significant obstacle for modern computational machines, especially when the gain in accuracy is so evident.

We should also emphasize that, in a purely random classification, we would have approximately $\approx (1/40)\%$ accuracy. In this sense, even when using the data directly, machine learning methods are more effective than simple random attempts. Consequently, by using recurrence microstates, we achieve a significantly higher efficiency gain.

Finally, let us consider the processing time of the machine learning algorithms, both when applied directly to the raw data and when applied to data transformed into recurrence microstates. Table 3 presents the processing times for the Random Forest algorithm, which, according to our study, achieved the best overall results. The complete results for all methods are provided in the appendices.

It can be observed that when using data of the recurrence microstate space, considering, for instance, $N=3$, which already provides a good level of accuracies, the CPU time required for data processing is reduced by at least a factor of 7 compared to using the raw data. In summary, the use of recurrence microstates not only improves the accuracy of machine learning methods but also significantly reduces processing time.

\begin{table*}
\centering
\begin{tabular}{|c|c|ccc|}\hline
\multicolumn{5}{|c|}{Mean Execution Time (s)} \\
\hline\hline
Model & Data & Logistic & Lorenz & Color Noises \\
\hline
\multirow[c]{4}{*}{Random Forest} & $N=2$ & 1.097 ± 0.082 & 1.348 ± 0.192 & 1.326 ± 0.488 \\
 & $N=3$ & 0.971 ± 0.026 & 3.934 ± 0.464 & 6.474 ± 4.563\\
 & $N=4$ & 6.800 ± 0.337 & 11.442 ± 0.390 & 39.929 ± 42.988\\
 & Raw Data & 6.971 ± 0.043 & 63.997 ± 1.847 & 75.863 ± 1.634 \\ 
\hline
\end{tabular}
\caption{Average execution times (in seconds) for the Random Forest algorithm, which showed the best trade-off between accuracy and computational cost. The table presents the mean and standard deviation of execution time for different feature extraction strategies ($N=2$, $N=3$, $N=4$) and for the raw time series data. The results are expressed as the mean ± standard deviation and were obtained from all 20 possible training and testing permutations across five distinct datasets. The evaluation was performed on three representative systems: a discrete chaotic map (Logistic), a continuous chaotic system (Lorenz), and stochastic colored noise signals (Color Noises).}
\label{tab:RF-times}
\end{table*}

\section{Discussions and conclusions}

In this study, we investigate the performance of different machine learning algorithms in classifying parameters of dynamical systems using recurrence microstate quantifiers as features. The features employed are recurrence microstate entropy, the optimal recurrence threshold, and the probability of occurrence of each microstate.

Our findings indicate that incorporating microstate-based features significantly enhances the classification accuracy of machine learning algorithms. Furthermore, we observe a positive correlation between microstate size and classification accuracy, suggesting that larger microstates capture richer dynamical information, thereby providing more discriminative features for the algorithms. This improvement in accuracy can be attributed to the increased informational content encoded by larger microstates, which enhances the ability of machine learning models to distinguish between different dynamical regimes. These results are associated with a characteristic of the recurrence microstate space, namely, its ability to evaluate a domain of points rather than just eventual relationships obtained from sequences of points in a time series.

However, the use of larger microstates is associated with increased computational time. This trade-off underscores the importance of selecting computationally efficient models, especially when dealing with large datasets. For instance, deep learning architectures such as multilayer perceptrons (MLPs) require a substantial number of parameters, making them less practical for large-scale applications without suitably pre-processing the data, as in the case of using recurrence microstate features. These insights contribute to the broader understanding of feature selection in dynamical systems classification.

Our results show that these features encode information about the generating process of the time series better than the raw data itself. When classifying the parameter that generates discrete, continuous, and noisy systems, we find that, for all five maps analyzed, the ten algorithms that achieve at least 10\% accuracy using microstates outperform those relying on the raw time series. For continuous systems (flows), there is greater variability in performance, but the algorithms that achieve acceptable accuracy with recurrence features perform up to 100 times better than when using raw data. A similar pattern is observed for noise classification: although the best models achieve accuracy between 40\% and 60\%, this is significantly higher than random guessing (2.5\%) and far superior to the raw data approach, which barely reaches 10\%.

Additionally, we observe that some algorithms consistently outperform others, particularly MLPs and Random Forest. Both exhibit comparable performance across all analyses. While MLPs achieve slightly higher accuracy in some cases, the difference is not substantial enough to justify the increased computational cost and complexity compared to Random Forest. The only notable exception is the Lorenz system, where MLP significantly outperforms Random Forest. Given the computational burden of large microstate-based feature sets, we recommend using Random Forest as the first approach due to its simplicity and efficiency. For large datasets, MLPs can lead to models with an overwhelming number of parameters, making them less practical unless a substantial accuracy improvement is required.

The relevance of this study extends to various fields dealing with complex time series data, such as neuroscience, climatology, and biomedical signal analysis. By demonstrating the applicability of recurrence quantifiers for dynamical system classification, this work contributes to understanding the underlying dynamics of highly complex and nonlinear phenomena and provides a foundation for developing new machine-learning-based methodologies for recurrence analysis.

We acknowledge that the approach presented relies on supervised learning, which requires pre-classified training data with known ground truth. While this is a limitation in many real-world applications where labeled datasets are scarce or costly, it is important to note that the recurrence entropy quantifier used here extracts subtle differences from time series that are visually indistinguishable. This capability enables a clear separation of dynamic states in both chaotic and stochastic systems, facilitating classification in a straightforward manner. Moreover, this property suggests potential applications beyond synthetic dynamical systems, including the classification or discrimination of real-world datasets such as climatic or neurological signals, which are currently the focus of ongoing research.

Finally, several points can be raised regarding the classification process discussed here. Our analysis is based on a transformation of spaces, treating the data in the microstate space as a complete set of information on which the classification process is carried out. Such a translation of the real characteristics of the data into the recurrence space may be affected by factors such as noise level, sampling adequacy, and dataset size. Another relevant point is that, in principle, the translation of information into the recurrence microstate space may be data-dependent, although this hypothesis is unlikely for dynamics exhibiting some degree of stationarity.

Future work will explore variations in recurrence parameters, incorporating additional complementary quantifiers, and applying this approach to larger experimental datasets. Additionally, integrating deep learning techniques may allow for a more sophisticated modeling of the temporal relationships in the data, enhancing the accuracies of dynamical systems classifications.


\section*{Acknowledgements}
This work is supported by the following Brazilian research agency:  Conselho Nacional de Desenvolvimento Cient\'ifico e Tecnol\'ogico (CNPq), Grants Nos. 308441/2021-4, 407072/2022-5, 305189/2022-0, 408254/2022-0 and 300064/2023-3. Co\-or\-de\-na\-ção de A\-per\-fei\-ço\-a\-men\-to de Pes\-so\-al de Nível Superior (CAPES), Grants Nos. 88881.895032/2023-01.

\section*{Statements and Declarations}

\textbf{Competing Interests}  
The authors declare that they have no competing interests, financial or non‑financial, that could have appeared to influence the work reported in this paper.

\textbf{Data Availability}  
The datasets generated during and/or analysed during the current study are available from the corresponding author on reasonable request.

\bibliography{example}


\begin{thebibliography}{34}
\ifx \bisbn   \undefined \def \bisbn  #1{ISBN #1}\fi
\ifx \binits  \undefined \def \binits#1{#1}\fi
\ifx \bauthor  \undefined \def \bauthor#1{#1}\fi
\ifx \batitle  \undefined \def \batitle#1{#1}\fi
\ifx \bjtitle  \undefined \def \bjtitle#1{#1}\fi
\ifx \bvolume  \undefined \def \bvolume#1{\textbf{#1}}\fi
\ifx \byear  \undefined \def \byear#1{#1}\fi
\ifx \bissue  \undefined \def \bissue#1{#1}\fi
\ifx \bfpage  \undefined \def \bfpage#1{#1}\fi
\ifx \blpage  \undefined \def \blpage #1{#1}\fi
\ifx \burl  \undefined \def \burl#1{\textsf{#1}}\fi
\ifx \doiurl  \undefined \def \doiurl#1{\url{https://doi.org/#1}}\fi
\ifx \betal  \undefined \def \betal{\textit{et al.}}\fi
\ifx \binstitute  \undefined \def \binstitute#1{#1}\fi
\ifx \binstitutionaled  \undefined \def \binstitutionaled#1{#1}\fi
\ifx \bctitle  \undefined \def \bctitle#1{#1}\fi
\ifx \beditor  \undefined \def \beditor#1{#1}\fi
\ifx \bpublisher  \undefined \def \bpublisher#1{#1}\fi
\ifx \bbtitle  \undefined \def \bbtitle#1{#1}\fi
\ifx \bedition  \undefined \def \bedition#1{#1}\fi
\ifx \bseriesno  \undefined \def \bseriesno#1{#1}\fi
\ifx \blocation  \undefined \def \blocation#1{#1}\fi
\ifx \bsertitle  \undefined \def \bsertitle#1{#1}\fi
\ifx \bsnm \undefined \def \bsnm#1{#1}\fi
\ifx \bsuffix \undefined \def \bsuffix#1{#1}\fi
\ifx \bparticle \undefined \def \bparticle#1{#1}\fi
\ifx \barticle \undefined \def \barticle#1{#1}\fi
\bibcommenthead
\ifx \bconfdate \undefined \def \bconfdate #1{#1}\fi
\ifx \botherref \undefined \def \botherref #1{#1}\fi
\ifx \url \undefined \def \url#1{\textsf{#1}}\fi
\ifx \bchapter \undefined \def \bchapter#1{#1}\fi
\ifx \bbook \undefined \def \bbook#1{#1}\fi
\ifx \bcomment \undefined \def \bcomment#1{#1}\fi
\ifx \oauthor \undefined \def \oauthor#1{#1}\fi
\ifx \citeauthoryear \undefined \def \citeauthoryear#1{#1}\fi
\ifx \endbibitem  \undefined \def \endbibitem {}\fi
\ifx \bconflocation  \undefined \def \bconflocation#1{#1}\fi
\ifx \arxivurl  \undefined \def \arxivurl#1{\textsf{#1}}\fi
\csname PreBibitemsHook\endcsname

\bibitem[\protect\citeauthoryear{Han et~al.}{2012}]{han2012data}
\begin{bbook}
\bauthor{\bsnm{Han}, \binits{J.}},
\bauthor{\bsnm{Kamber}, \binits{M.}},
\bauthor{\bsnm{Pei}, \binits{J.}}:
\bbtitle{Data Mining: Concepts and Techniques},
\bedition{3rd} edn.
\bpublisher{Morgan Kaufmann},
\blocation{San Francisco}
(\byear{2012})
\end{bbook}
\endbibitem

\bibitem[\protect\citeauthoryear{Géron}{2019}]{maosaobra}
\begin{bbook}
\bauthor{\bsnm{Géron}, \binits{A.}}:
\bbtitle{Hands-on Machine Learning with Scikit-Learn, Keras, and TensorFlow: Concepts, Tools, and Techniques to Build Intelligent Systems}.
\bpublisher{O'Reilly Media},
\blocation{Sebastopol, CA}
(\byear{2019})
\end{bbook}
\endbibitem

\bibitem[\protect\citeauthoryear{Haykin}{2009}]{haykin2009neural}
\begin{bbook}
\bauthor{\bsnm{Haykin}, \binits{S.}}:
\bbtitle{Neural Networks and Learning Machines},
\bedition{3rd} edn.
\bpublisher{Prentice Hall},
\blocation{Upper Saddle River, NJ}
(\byear{2009})
\end{bbook}
\endbibitem

\bibitem[\protect\citeauthoryear{Pathria and Beale}{2011}]{pathria2016statistical}
\begin{bbook}
\bauthor{\bsnm{Pathria}, \binits{R.K.}},
\bauthor{\bsnm{Beale}, \binits{P.D.}}:
\bbtitle{Statistical Mechanics},
\bedition{3rd} edn.
\bpublisher{Butterworth-Heinemann},
\blocation{Oxford}
(\byear{2011})
\end{bbook}
\endbibitem

\bibitem[\protect\citeauthoryear{Strogatz}{2001}]{STROGATZ2001}
\begin{barticle}
\bauthor{\bsnm{Strogatz}, \binits{S.H.}}:
\batitle{Exploring complex networks}.
\bjtitle{Nature}
\bvolume{410},
\bfpage{268}--\blpage{276}
(\byear{2001})
\doiurl{10.1038/35065725}
\end{barticle}
\endbibitem

\bibitem[\protect\citeauthoryear{Velliangiri et~al.}{2019}]{velliangiri2019review}
\begin{barticle}
\bauthor{\bsnm{Velliangiri}, \binits{S.}},
\bauthor{\bsnm{Alagumuthukrishnan}, \binits{S.}},
\bauthor{\bsnm{Iwin}, \binits{S.T.J.}}:
\batitle{A review of dimensionality reduction techniques for efficient computation}.
\bjtitle{Procedia Computer Science}
\bvolume{165},
\bfpage{104}--\blpage{111}
(\byear{2019})
\doiurl{10.1016/j.procs.2020.01.079}
\end{barticle}
\endbibitem

\bibitem[\protect\citeauthoryear{Jolliffe and Cadima}{2016}]{jolliffe2016principal}
\begin{barticle}
\bauthor{\bsnm{Jolliffe}, \binits{I.T.}},
\bauthor{\bsnm{Cadima}, \binits{J.}}:
\batitle{Principal component analysis: a review and recent developments}.
\bjtitle{Philosophical Transactions of the Royal Society A: Mathematical, Physical and Engineering Sciences}
\bvolume{374},
\bfpage{20150202}
(\byear{2016})
\doiurl{10.1098/rsta.2015.0202}
\end{barticle}
\endbibitem

\bibitem[\protect\citeauthoryear{McInnes and Healy}{2018}]{mcinnes2018umap}
\begin{botherref}
\oauthor{\bsnm{McInnes}, \binits{L.}},
\oauthor{\bsnm{Healy}, \binits{J.}}:
UMAP: Uniform Manifold Approximation and Projection for Dimension Reduction
(2018).
\doiurl{10.48550/arXiv.1802.03426}
\end{botherref}
\endbibitem

\bibitem[\protect\citeauthoryear{Chandrashekar and Sahin}{2014}]{chandrashekar2014survey}
\begin{barticle}
\bauthor{\bsnm{Chandrashekar}, \binits{G.}},
\bauthor{\bsnm{Sahin}, \binits{F.}}:
\batitle{A survey on feature selection methods}.
\bjtitle{Computers \& Electrical Engineering}
\bvolume{40}(\bissue{1}),
\bfpage{16}--\blpage{28}
(\byear{2014})
\doiurl{10.1016/j.compeleceng.2013.11.024}
\end{barticle}
\endbibitem

\bibitem[\protect\citeauthoryear{Bhagoji et~al.}{2018}]{MLDataTransformations}
\begin{bchapter}
\bauthor{\bsnm{Bhagoji}, \binits{A.N.}},
\bauthor{\bsnm{Cullina}, \binits{D.}},
\bauthor{\bsnm{Sitawarin}, \binits{C.}},
\bauthor{\bsnm{Mittal}, \binits{P.}}:
\bctitle{Enhancing robustness of machine learning systems via data transformations}.
In: \bbtitle{2018 52nd Annual Conference on Information Sciences and Systems (CISS)},
pp. \bfpage{1}--\blpage{5}
(\byear{2018}).
\doiurl{10.1109/CISS.2018.8362326}
\end{bchapter}
\endbibitem

\bibitem[\protect\citeauthoryear{Jia et~al.}{2022}]{jia2022feature}
\begin{botherref}
\oauthor{\bsnm{Jia}, \binits{W.}},
\oauthor{\bsnm{Sun}, \binits{M.}},
\oauthor{\bsnm{Lian}, \binits{J.}},
\oauthor{\bsnm{Hou}, \binits{S.}}:
Feature dimensionality reduction: a review.
Complex \& Intelligent Systems
\textbf{8}
(2022)
\doiurl{10.1007/s40747-021-00637-x}
\end{botherref}
\endbibitem

\bibitem[\protect\citeauthoryear{Marwan et~al.}{2007}]{Marwan2007}
\begin{barticle}
\bauthor{\bsnm{Marwan}, \binits{N.}},
\bauthor{\bsnm{Romano}, \binits{M.C.}},
\bauthor{\bsnm{Thiel}, \binits{M.}},
\bauthor{\bsnm{Kurths}, \binits{J.}}:
\batitle{Recurrence plots for the analysis of complex systems}.
\bjtitle{Physics Reports}
\bvolume{438}(\bissue{5}),
\bfpage{237}--\blpage{329}
(\byear{2007})
\doiurl{10.1016/j.physrep.2006.11.001}
\end{barticle}
\endbibitem

\bibitem[\protect\citeauthoryear{Poincar{\'e}}{1890}]{Poincare1890}
\begin{bbook}
\bauthor{\bsnm{Poincar{\'e}}, \binits{H.}}:
\bbtitle{Sur Le Probleme des Trois Corps et les Equations de la Dynamique}.
\bsertitle{Acta Mathematica}.
\bpublisher{F. \& G. Beijer},
\blocation{Stockholm}
(\byear{1890})
\end{bbook}
\endbibitem

\bibitem[\protect\citeauthoryear{Eckmann et~al.}{1987}]{Eckmann1987}
\begin{barticle}
\bauthor{\bsnm{Eckmann}, \binits{J.-P.}},
\bauthor{\bsnm{Kamphorst}, \binits{S.O.}},
\bauthor{\bsnm{Ruelle}, \binits{D.}}:
\batitle{Recurrence plots of dynamical systems}.
\bjtitle{Europhysics Letters}
\bvolume{4}(\bissue{9}),
\bfpage{973}
(\byear{1987})
\doiurl{10.1209/0295-5075/4/9/004}
\end{barticle}
\endbibitem

\bibitem[\protect\citeauthoryear{Zbilut and Webber}{2007}]{zibilut2007}
\begin{barticle}
\bauthor{\bsnm{Zbilut}, \binits{J.P.}},
\bauthor{\bsnm{Webber}, \binits{C.L.}}:
\batitle{Recurrence quantification analysis: Introduction and historical context}.
\bjtitle{International Journal of Bifurcation and Chaos}
\bvolume{17}(\bissue{10}),
\bfpage{3477}--\blpage{3481}
(\byear{2007})
\doiurl{10.1142/S0218127407019238}
\end{barticle}
\endbibitem

\bibitem[\protect\citeauthoryear{Marwan and Kraemer}{2023}]{Marwan2023}
\begin{barticle}
\bauthor{\bsnm{Marwan}, \binits{N.}},
\bauthor{\bsnm{Kraemer}, \binits{K.H.}}:
\batitle{Trends in recurrence analysis of dynamical systems}.
\bjtitle{The European Physical Journal Special Topics}
\bvolume{232}(\bissue{1}),
\bfpage{5}--\blpage{27}
(\byear{2023})
\doiurl{10.1140/epjs/s11734-022-00739-8}
\end{barticle}
\endbibitem

\bibitem[\protect\citeauthoryear{Corso et~al.}{2018}]{Corso2018}
\begin{botherref}
\oauthor{\bsnm{Corso}, \binits{G.}},
\oauthor{\bsnm{Prado}, \binits{T.L.}},
\oauthor{\bsnm{Lima}, \binits{G.Z.S.}},
\oauthor{\bsnm{Kurths}, \binits{J.}},
\oauthor{\bsnm{Lopes}, \binits{S.R.}}:
Quantifying entropy using recurrence matrix microstates.
Chaos: An Interdisciplinary Journal of Nonlinear Science
\textbf{28}(8)
(2018)
\doiurl{10.1063/1.5042026}
\end{botherref}
\endbibitem

\bibitem[\protect\citeauthoryear{Prado et~al.}{2020}]{Prado2020}
\begin{botherref}
\oauthor{\bsnm{Prado}, \binits{T.L.}},
\oauthor{\bsnm{Corso}, \binits{G.}},
\oauthor{\bsnm{Lima}, \binits{G.Z.S.S.}},
\oauthor{\bsnm{Budzinski}, \binits{R.C.}},
\oauthor{\bsnm{Boaretto}, \binits{B.R.R.}},
\oauthor{\bsnm{Ferrari}, \binits{F.A.S.}},
\oauthor{\bsnm{Macau}, \binits{E.E.N.}},
\oauthor{\bsnm{Lopes}, \binits{S.R.}}:
Maximum entropy principle in recurrence plot analysis on stochastic and chaotic systems.
Chaos: An Interdisciplinary Journal of Nonlinear Science
\textbf{30}(4)
(2020)
\doiurl{10.1063/1.5125921}
\end{botherref}
\endbibitem

\bibitem[\protect\citeauthoryear{Corso et~al.}{2021}]{corso2021maximum}
\begin{barticle}
\bauthor{\bsnm{Corso}, \binits{G.}},
\bauthor{\bsnm{Lima}, \binits{G.Z.S.}},
\bauthor{\bsnm{Lopes}, \binits{S.R.}},
\bauthor{\bsnm{Prado}, \binits{T.L.}},
\bauthor{\bsnm{Correa}, \binits{M.A.}},
\bauthor{\bsnm{Bohn}, \binits{F.}}:
\batitle{Maximum entropy in the dimensional transition of the magnetic domain wall dynamics}.
\bjtitle{Physica A: Statistical Mechanics and its Applications}
\bvolume{568},
\bfpage{125730}
(\byear{2021})
\doiurl{10.1016/j.physa.2021.125730}
\end{barticle}
\endbibitem

\bibitem[\protect\citeauthoryear{Boaretto et~al.}{2024}]{boaretto2024use}
\begin{barticle}
\bauthor{\bsnm{Boaretto}, \binits{B.R.R.}},
\bauthor{\bsnm{Andreani}, \binits{A.C.}},
\bauthor{\bsnm{Lopes}, \binits{S.R.}},
\bauthor{\bsnm{Prado}, \binits{T.L.}},
\bauthor{\bsnm{Macau}, \binits{E.E.N.}}:
\batitle{The use of entropy of recurrence microstates and artificial intelligence to detect cardiac arrhythmia in ecg records}.
\bjtitle{Applied Mathematics and Computation}
\bvolume{475},
\bfpage{128738}
(\byear{2024})
\doiurl{10.1016/j.amc.2024.128738}
\end{barticle}
\endbibitem

\bibitem[\protect\citeauthoryear{Ferré et~al.}{2024}]{ferre2024cycling}
\begin{barticle}
\bauthor{\bsnm{Ferré}, \binits{I.B.S.}},
\bauthor{\bsnm{Corso}, \binits{G.}},
\bauthor{\bsnm{Lima}, \binits{G.Z.S.}},
\bauthor{\bsnm{Lopes}, \binits{S.R.}},
\bauthor{\bsnm{Leocadio-Miguel}, \binits{M.A.}},
\bauthor{\bsnm{França}, \binits{L.G.S.}},
\bauthor{\bsnm{Prado}, \binits{T.L.}},
\bauthor{\bsnm{Araújo}, \binits{J.F.}}:
\batitle{Cycling reduces the entropy of neuronal activity in the human adult cortex}.
\bjtitle{PLOS ONE}
(\byear{2024})
\doiurl{10.1371/journal.pone.0298703}
\end{barticle}
\endbibitem

\bibitem[\protect\citeauthoryear{Prado et~al.}{2021}]{Prado2021}
\begin{barticle}
\bauthor{\bsnm{Prado}, \binits{T.d.L.}},
\bauthor{\bsnm{Macau}, \binits{E.E.N.}},
\bauthor{\bsnm{Lopes}, \binits{S.R.}}:
\batitle{Detection of data corruption in stationary time series using recurrence microstates probabilities}.
\bjtitle{The European Physical Journal Special Topics}
\bvolume{230}(\bissue{14–15}),
\bfpage{2737}--\blpage{2744}
(\byear{2021})
\doiurl{10.1140/epjs/s11734-021-00169-y}
\end{barticle}
\endbibitem

\bibitem[\protect\citeauthoryear{Spezzatto et~al.}{2024}]{spezzatto2024recurrence}
\begin{barticle}
\bauthor{\bsnm{Spezzatto}, \binits{G.S.}},
\bauthor{\bsnm{Flauzino}, \binits{J.V.V.}},
\bauthor{\bsnm{Corso}, \binits{G.}},
\bauthor{\bsnm{Boaretto}, \binits{B.R.R.}},
\bauthor{\bsnm{Macau}, \binits{E.E.N.}},
\bauthor{\bsnm{Prado}, \binits{T.L.}},
\bauthor{\bsnm{Lopes}, \binits{S.R.}}:
\batitle{Recurrence microstates for machine learning classification}.
\bjtitle{Chaos}
\bvolume{34},
\bfpage{073140}
(\byear{2024})
\doiurl{10.1063/5.0203801}
\end{barticle}
\endbibitem

\bibitem[\protect\citeauthoryear{Pedregosa et~al.}{2011}]{scikit-learn}
\begin{barticle}
\bauthor{\bsnm{Pedregosa}, \binits{F.}},
\bauthor{\bsnm{Varoquaux}, \binits{G.}},
\bauthor{\bsnm{Gramfort}, \binits{A.}},
\bauthor{\bsnm{Michel}, \binits{V.}},
\bauthor{\bsnm{Thirion}, \binits{B.}},
\bauthor{\bsnm{Grisel}, \binits{O.}},
\bauthor{\bsnm{Blondel}, \binits{M.}},
\bauthor{\bsnm{Prettenhofer}, \binits{P.}},
\bauthor{\bsnm{Weiss}, \binits{R.}},
\bauthor{\bsnm{Dubourg}, \binits{V.}},
\bauthor{\bsnm{Vanderplas}, \binits{J.}},
\bauthor{\bsnm{Passos}, \binits{A.}},
\bauthor{\bsnm{Cournapeau}, \binits{D.}},
\bauthor{\bsnm{Brucher}, \binits{M.}},
\bauthor{\bsnm{Perrot}, \binits{M.}},
\bauthor{\bsnm{Duchesnay}, \binits{E.}}:
\batitle{Scikit-learn: Machine learning in {P}ython}.
\bjtitle{Journal of Machine Learning Research}
\bvolume{12},
\bfpage{2825}--\blpage{2830}
(\byear{2011})
\end{barticle}
\endbibitem

\bibitem[\protect\citeauthoryear{Escovedo and Koshiyama}{2020}]{introdatasci}
\begin{bbook}
\bauthor{\bsnm{Escovedo}, \binits{T.}},
\bauthor{\bsnm{Koshiyama}, \binits{A.}}:
\bbtitle{Introdu{\c{c}}{\~a}o a Data Science: Algoritmos de Machine Learning e M{\'e}todos de an{\'a}lise}.
\bpublisher{Casa do C{\'o}digo},
\blocation{São Paulo}
(\byear{2020})
\end{bbook}
\endbibitem

\bibitem[\protect\citeauthoryear{Hilborn}{2004}]{ChaosandNonlinear}
\begin{bbook}
\bauthor{\bsnm{Hilborn}, \binits{R.C.}}:
\bbtitle{Chaos and Nonlinear Dynamics: An Introduction for Scientists and Engineers}.
\bpublisher{Oxford University Press},
\blocation{New York}
(\byear{2004})
\end{bbook}
\endbibitem

\bibitem[\protect\citeauthoryear{Weissman}{1988}]{Weissman1988}
\begin{barticle}
\bauthor{\bsnm{Weissman}, \binits{M.B.}}:
\batitle{1/f noise and other slow, nonexponential kinetics}.
\bjtitle{Reviews of Modern Physics}
\bvolume{60}(\bissue{2}),
\bfpage{537}--\blpage{571}
(\byear{1988})
\doiurl{10.1103/RevModPhys.60.537}
\end{barticle}
\endbibitem

\bibitem[\protect\citeauthoryear{McCulloch and Pitts}{1943}]{mcculloch1943logical}
\begin{barticle}
\bauthor{\bsnm{McCulloch}, \binits{W.S.}},
\bauthor{\bsnm{Pitts}, \binits{W.}}:
\batitle{A logical calculus of the ideas immanent in nervous activity}.
\bjtitle{The bulletin of mathematical biophysics}
\bvolume{5},
\bfpage{115}--\blpage{133}
(\byear{1943})
\end{barticle}
\endbibitem

\bibitem[\protect\citeauthoryear{Chollet et~al.}{2015}]{chollet2015keras}
\begin{botherref}
\oauthor{\bsnm{Chollet}, \binits{F.}}, et al.:
Keras.
\url{https://keras.io}
(2015)
\end{botherref}
\endbibitem

\bibitem[\protect\citeauthoryear{Thieler et~al.}{2016}]{thieler2016robper}
\begin{barticle}
\bauthor{\bsnm{Thieler}, \binits{A.M.}},
\bauthor{\bsnm{Fried}, \binits{R.}},
\bauthor{\bsnm{Rathjens}, \binits{J.}}:
\batitle{Robper: An {R} package to calculate periodograms for light curves based on robust regression}.
\bjtitle{Journal of Statistical Software}
\bvolume{69}(\bissue{9}),
\bfpage{1}--\blpage{36}
(\byear{2016})
\doiurl{10.18637/jss.v069.i09}
\end{barticle}
\endbibitem

\bibitem[\protect\citeauthoryear{{R Core Team}}{2024}]{R2024}
\begin{botherref}
\oauthor{\bsnm{{R Core Team}}}:
R: A Language and Environment for Statistical Computing.
R Foundation for Statistical Computing,
(2024).
R Foundation for Statistical Computing.
\url{https://www.R-project.org/}
\end{botherref}
\endbibitem

\bibitem[\protect\citeauthoryear{Bezanson et~al.}{2017}]{Julia17}
\begin{barticle}
\bauthor{\bsnm{Bezanson}, \binits{J.}},
\bauthor{\bsnm{Edelman}, \binits{A.}},
\bauthor{\bsnm{Karpinski}, \binits{S.}},
\bauthor{\bsnm{Shah}, \binits{V.B.}}:
\batitle{Julia: A fresh approach to numerical computing}.
\bjtitle{SIAM review}
\bvolume{59}(\bissue{1}),
\bfpage{65}--\blpage{98}
(\byear{2017})
\end{barticle}
\endbibitem

\bibitem[\protect\citeauthoryear{Wickramasinghe and Kalutarage}{2021}]{wickramasinghe2021naive}
\begin{barticle}
\bauthor{\bsnm{Wickramasinghe}, \binits{I.}},
\bauthor{\bsnm{Kalutarage}, \binits{H.}}:
\batitle{Naive bayes: applications, variations and vulnerabilities: a review of literature with code snippets for implementation}.
\bjtitle{Soft Computing}
\bvolume{25}(\bissue{3}),
\bfpage{2277}--\blpage{2293}
(\byear{2021})
\doiurl{10.1007/s00500-020-05297-6}
\end{barticle}
\endbibitem

\bibitem[\protect\citeauthoryear{Cover}{1965}]{cover1965geometrical}
\begin{barticle}
\bauthor{\bsnm{Cover}, \binits{T.M.}}:
\batitle{Geometrical and statistical properties of systems of linear inequalities with applications in pattern recognition}.
\bjtitle{IEEE Transactions on Electronic Computers}
\bvolume{EC-14}(\bissue{3}),
\bfpage{326}--\blpage{334}
(\byear{1965})
\doiurl{10.1109/PGEC.1965.264137}
\end{barticle}
\endbibitem

\end{thebibliography}

\newpage
\begin{appendices}

\section{Mean execution time of ML routines applied in recurrence microstate of discrete dynamical systems data}\label{secA1}

Here, we report the mean CPU processing times for all machine learning methods employed in this study and using nonlinear dynamics of discrete data, considering the scenario in which the input data were pre-processed using recurrence microstates. Table \ref{tab:average-time-maps} presents the results for three different microstate sizes, $N=2$,$N=3$, and $N=4$.
\begin{table*}[hbp]
\resizebox{\textwidth}{!}{\begin{tabular}{|c|c|ccccc|}
\hline
 &  &  \multicolumn{5}{c|}{Mean Execution Time (s)}  \\
Model & $N$ & BetaX & Gauss & Henon & Ikeda & Logistic \\
\hline
\hline
\multirow[m]{3}{*}{Bernoulli NB} & 2 & 0.007 ± 0.002 & 0.004 ± 0.001 & 0.007 ± 0.001 & 0.003 ± 0.000 & 0.014 ± 0.004  \\
 & 3 & 0.022 ± 0.002 & 0.018 ± 0.001 & 0.017 ± 0.001 & 0.015 ± 0.000 & 0.025 ± 0.002  \\
 & 4 & 5.161 ± 0.381 & 4.491 ± 0.438 & 2.971 ± 0.219 & 4.335 ± 0.881 & 3.250 ± 0.497  \\
\hline
\multirow[m]{3}{*}{Decision Tree} & 2 & 0.055 ± 0.009 & 0.052 ± 0.021 & 0.053 ± 0.007 & 0.041 ± 0.007 & 0.067 ± 0.007 \\
 & 3 & 0.626 ± 0.041 & 0.292 ± 0.028 & 0.220 ± 0.011 & 0.565 ± 0.036 & 0.346 ± 0.004  \\
 & 4 & 38.994 ± 2.010 & 8.051 ± 0.603 & 5.480 ± 0.416 & 29.351 ± 1.138 & 8.868 ± 1.415 \\
\hline
\multirow[m]{3}{*}{Gaussian NB} & 2 & 0.017 ± 0.002 & 0.013 ± 0.001 & 0.017 ± 0.003 & 0.010 ± 0.000 & 0.025 ± 0.009 \\
 & 3 & 0.220 ± 0.033 & 0.125 ± 0.016 & 0.109 ± 0.010 & 0.107 ± 0.003 & 0.258 ± 0.040  \\
 & 4 & 35.951 ± 0.566 & 53.523 ± 0.735 & 36.153 ± 0.441 & 37.006 ± 0.773 & 35.664 ± 0.538  \\
\hline
\multirow[m]{3}{*}{Gradient Boosting} & 2 & 44.096 ± 0.528 & 26.636 ± 0.931 & 23.811 ± 1.071 & 25.833 ± 0.721 & 40.631 ± 0.859 \\
 & 3 & 402.574 ± 12.849 & 178.393 ± 0.533 & 196.466 ± 0.275 & 347.399 ± 1.891 & 208.609 ± 1.267 \\
 & 4 & 71.321 ± 2.662 & 56.189 ± 1.215 & 46.232 ± 0.928 & 58.876 ± 0.536 & 40.393 ± 1.810  \\
\hline
\multirow[m]{3}{*}{KNN} & 2 & 0.252 ± 0.012 & 0.183 ± 0.009 & 0.218 ± 0.036 & 0.038 ± 0.006 & 0.323 ± 0.082  \\
 & 3 & 0.240 ± 0.046 & 0.104 ± 0.018 & 0.090 ± 0.049 & 0.059 ± 0.009 & 0.201 ± 0.019  \\
 & 4 & 15.827 ± 0.026 & 17.459 ± 0.015 & 16.036 ± 0.046 & 15.719 ± 0.002 & 15.921 ± 0.042 \\
\hline
\multirow[m]{3}{*}{Linear SVC} & 2 & 1.014 ± 0.040 & 0.353 ± 0.015 & 0.689 ± 0.008 & 0.477 ± 0.009 & 0.679 ± 0.041  \\
 & 3 & 50.865 ± 1.435 & 4.626 ± 0.183 & 8.066 ± 0.260 & 18.311 ± 0.338 & 5.831 ± 0.178 \\
 & 4 & 4185.380 ± 4.231 & 198.827 ± 2.399 & 249.403 ± 7.436 & 1323.546 ± 3.359 & 170.252 ± 3.704  \\
\hline
\multirow[m]{3}{*}{Logistic Regression} & 2 & 0.124 ± 0.008 & 0.215 ± 0.022 & 0.143 ± 0.006 & 0.269 ± 0.070 & 0.490 ± 0.105  \\
 & 3 & 1.030 ± 0.296 & 1.041 ± 0.164 & 0.896 ± 0.105 & 1.081 ± 0.262 & 1.760 ± 0.269 \\
 & 4 & 184.663 ± 2.109 & 226.645 ± 1.499 & 184.072 ± 2.722 & 234.911 ± 3.303 & 185.079 ± 2.155  \\
\hline
\multirow[m]{3}{*}{MLP} & 2 & 10.938 ± 0.164 & 5.670 ± 0.156 & 6.165 ± 0.893 & 5.210 ± 0.407 & 11.582 ± 1.392  \\
 & 3 & 7.014 ± 0.184 & 5.509 ± 0.088 & 5.509 ± 0.122 & 5.441 ± 0.135 & 6.588 ± 0.301  \\
 & 4 & 409.604 ± 10.810 & 505.888 ± 22.794 & 432.969 ± 0.024 & 422.836 ± 0.250 & 433.998 ± 0.710\\
\hline
\multirow[m]{3}{*}{Random Forest} & 2 & 1.024 ± 0.066 & 0.672 ± 0.015 & 0.989 ± 0.030 & 0.597 ± 0.008 & 1.097 ± 0.082  \\
 & 3 & 1.825 ± 0.022 & 0.962 ± 0.010 & 0.898 ± 0.005 & 1.493 ± 0.005 & 0.971 ± 0.026  \\
 & 4 & 13.132 ± 0.272 & 7.971 ± 0.044 & 9.492 ± 2.708 & 10.726 ± 0.199 & 6.800 ± 0.337  \\
\hline
\multirow[m]{3}{*}{SVC} & 2 & 0.613 ± 0.021 & 0.359 ± 0.016 & 0.505 ± 0.003 & 0.316 ± 0.001 & 0.707 ± 0.061 \\
 & 3 & 1.586 ± 0.099 & 1.333 ± 0.020 & 1.093 ± 0.006 & 1.124 ± 0.005 & 1.549 ± 0.132 \\
 & 4 & 763.095 ± 8.471 & 997.667 ± 4.794 & 693.417 ± 7.129 & 645.754 ± 5.602 & 677.828 ± 4.960 \\
\hline
\end{tabular}}
\caption{Average execution times (in seconds) for different machine learning models, considering all 20 possible training and testing permutations across 5 distinct datasets. The values represent the mean and standard deviation of execution time for each model and microstate size $N$. The dataset size is $1600 \times 2^{(N \times N)}$, evaluated on the discrete dynamical systems BetaX, Gauss, Henon, Ikeda and Logistic} \label{tab:average-time-maps}
\end{table*}

\section{Mean execution time of ML routines applied in recurrence microstate of nonlinear continuous dynamical systems data}\label{secA2}

Here, we report the mean CPU processing times for all machine learning methods employed in this study and using nonlinear dynamical continuous systems data, considering the scenario in which the input data were pre-processed using recurrence microstates. Table \ref{tab:average-time-flows} presents the results for three different microstate sizes, $N=2$,$N=3$, and $N=4$.
\begin{table*}[btp]
\resizebox{\textwidth}{!}{
\begin{tabular}{|c|ccc|ccc|}
\hline
 &  \multicolumn{6}{c|}{Mean Execution Time (s)}  \\
 \hline
 & \multicolumn{3}{c|}{Lorenz} & \multicolumn{3}{c|}{Rossler} \\
Model & $N=2$ & $N=3$ & $N=4$ & $N=2$ & $N=3$ & $N=4$ \\
\hline
\hline
Bernoulli NB & 0.007 ± 0.002 & 0.030 ± 0.005 & 4.425 ± 0.660 & 0.006 ± 0.000 & 0.024 ± 0.000 & 3.923 ± 0.479 \\
Decision Tree & 0.062 ± 0.003 & 1.176 ± 0.154 & 24.797 ± 0.529 & 0.070 ± 0.009 & 1.116 ± 0.161 & 21.514 ± 1.258 \\
Gaussian NB & 0.021 ± 0.005 & 0.135 ± 0.009 & 34.692 ± 1.391 & 0.021 ± 0.002 & 0.185 ± 0.023 & 33.685 ± 2.012 \\
Gradient Boosting & 10.957 ± 0.438 & 31.352 ± 1.354 & 105.008 ± 4.060 & 10.502 ± 0.959 & 36.859 ± 0.715 & 87.941 ± 2.863 \\
KNN & 0.052 ± 0.074 & 0.115 ± 0.029 & 13.928 ± 0.283 & 0.028 ± 0.021 & 0.104 ± 0.005 & 15.756 ± 1.927 \\
Linear SVC & 0.069 ± 0.003 & 1.519 ± 0.170 & 3568.509 ± 57.397 & 0.089 ± 0.016 & 1.953 ± 0.220 & 2505.845 ± 49.497 \\
Logistic Regression & 0.062 ± 0.003 & 1.299 ± 0.764 & 59.506 ± 9.814 & 0.066 ± 0.003 & 0.911 ± 0.292 & 165.866 ± 4.570 \\
MLP & 16.809 ± 3.027 & 16.458 ± 0.215 & 322.543 ± 7.237 & 17.051 ± 1.194 & 16.951 ± 0.510 & 323.900 ± 3.201 \\
Random Forest & 1.348 ± 0.192 & 3.934 ± 0.464 & 11.442 ± 0.390 & 1.171 ± 0.157 & 3.931 ± 0.401 & 9.903 ± 0.227 \\
SVC & 0.555 ± 0.117 & 2.256 ± 0.342 & 655.843 ± 4.973 & 0.660 ± 0.159 & 2.035 ± 0.373 & 686.503 ± 23.068 \\
\hline
\end{tabular}
}
\caption{Average execution times (in seconds) for different machine learning models, considering all 20 possible training and testing permutations across 5 distinct datasets. The values represent the mean and standard deviation of execution time for each model and microstate size $N$. The dataset size is $1600 \times 2^{(N \times N)}$, evaluated on the continuous dynamical systems Lorenz and Rössler.}
\label{tab:average-time-flows}
\end{table*}

\newpage

\section{Mean execution time of ML routines applied direct in the raw data}\label{secA3}

Here, we report the mean CPU processing times for all machine learning methods employed in this study using nonlinear discrete and continuo dynamical systems   data, when the input data is the raw data obtained from the systems, without any pre-processed recurrence microstates. Table \ref{tab:raw_data}. 
\begin{table*}
\resizebox{1.0\textwidth}{!}{%
\begin{tabular}{|c|ccccc|cc|}
\hline
\multicolumn{1}{|c|}{} & \multicolumn{7}{c|}{Mean Execution Time (s)} \\
\multicolumn{1}{|c|}{} & \multicolumn{5}{c|}{Maps} & \multicolumn{2}{c|}{Flows} \\
Model & BetaX & Gauss & Henon & Ikeda & Logistic & Lorenz & Rossler \\
\hline\hline
Bernoulli NB & 0.024 ± 0.001 & 0.028 ± 0.001 & 0.026 ± 0.000 & 0.025 ± 0.000 & 0.024 ± 0.000 & 0.219 ± 0.037 & 0.225 ± 0.034 \\
Decision Tree & 6.032 ± 0.268 & 3.471 ± 0.316 & 3.032 ± 0.142 & 2.244 ± 0.135 & 3.378 ± 0.199 & 82.506 ± 2.305 & 17.006 ± 2.100 \\
Gaussian NB & 0.308 ± 0.009 & 0.371 ± 0.030 & 0.337 ± 0.003 & 0.318 ± 0.002 & 0.298 ± 0.007 & 1.615 ± 0.014 & 1.683 ± 0.107 \\
Gradient Boosting & 20.097 ± 0.045 & 18.841 ± 0.114 & 19.531 ± 0.083 & 17.953 ± 0.135 & 18.025 ± 0.035 & 121.315 ± 0.801 & 118.164 ± 0.647 \\
KNN & 0.108 ± 0.008 & 0.148 ± 0.002 & 0.101 ± 0.008 & 0.135 ± 0.032 & 0.095 ± 0.015 & 0.701 ± 0.145 & 0.696 ± 0.157 \\
Linear SVC & 4.890 ± 0.039 & 28.358 ± 0.282 & 25.730 ± 0.282 & 38.457 ± 0.249 & 37.412 ± 2.202 & 36.134 ± 0.878 & 178.266 ± 38.141 \\
Logistic Regression & 1.829 ± 0.133 & 2.044 ± 0.002 & 2.071 ± 0.005 & 1.927 ± 0.170 & 1.731 ± 0.116 & 1.072 ± 0.151 & 14.964 ± 0.481 \\
MLP & 6.467 ± 0.196 & 6.191 ± 0.047 & 6.229 ± 0.103 & 6.291 ± 0.092 & 6.013 ± 0.029 & 34.265 ± 1.392 & 34.667 ± 1.400 \\
Random Forest & 13.703 ± 0.184 & 7.180 ± 0.022 & 6.456 ± 0.593 & 4.514 ± 0.019 & 6.971 ± 0.043 & 63.997 ± 1.847 & 20.979 ± 0.635 \\
SVC & 2.723 ± 0.001 & 2.376 ± 0.075 & 2.878 ± 0.011 & 2.645 ± 0.083 & 2.778 ± 0.005 & 16.049 ± 0.093 & 15.609 ± 0.219 \\
\hline
\end{tabular}}
\caption{Average execution times (in seconds) for the different machine learning models using the raw time series data. 
The results consider all 20 possible training and testing permutations across 5 distinct datasets. 
The values represent the mean and standard deviation of execution time for each model. 
The evaluation was performed separately on discrete systems (BetaX, Gauss, Henon, Ikeda, Logistic) and continuous systems (Lorenz and Rossler).}
\label{tab:raw_data}
\end{table*}

\section{Mean execution time of ML routines applied in colored noises data}\label{secA4}

Here, we report the mean CPU processing times for all machine learning methods employed in this study using colored noises data. Table \ref{tab:colored_noises} presents the results for three different microstates sizes, $N=2$, $N=3$, $N=4$, as well as the results for the raw data, without the pre-processing.

\begin{table*}[btp]
\centering
\begin{tabular}{|c|cccc|}
\hline
 & \multicolumn{4}{c|}{Mean Execution Time (s)} \\
Model & $N=2$ & $N=3$ & $N=4$ & Raw Data \\
\hline
\hline
BernoulliNB & 0.008 ± 0.014 & 0.042 ± 0.045 & 4.510 ± 0.752 & 0.268 ± 0.108 \\
DecisionTreeClassifier & 0.085 ± 0.031 & 2.294 ± 1.984 & 178.393 ± 251.059 & 113.527 ± 12.999 \\
GaussianNB & 0.016 ± 0.003 & 0.153 ± 0.023 & 35.617 ± 2.176 & 3.174 ± 1.059 \\
GradientBoostingClassifier & 10.689 ± 0.662 & 37.962 ± 2.546 & 174.180 ± 4.977 & 160.567 ± 1.336 \\
KNeighborsClassifier & 0.024 ± 0.009 & 0.122 ± 0.036 & 15.457 ± 0.163 & 1.034 ± 0.204 \\
LinearSVC & 0.114 ± 0.012 & 3.564 ± 0.355 & 7914.803 ± 1038.042 & 420.400 ± 9.813 \\
LogisticRegression & 0.309 ± 0.276 & 6.493 ± 1.641 & 173.822 ± 4.289 & 28.690 ± 0.577 \\
MLP & 17.970 ± 0.591 & 20.141 ± 0.758 & 339.688 ± 35.348 & 52.835 ± 4.484 \\
RandomForestClassifier & 1.326 ± 0.489 & 6.474 ± 4.564 & 39.919 ± 42.988 & 75.863 ± 1.635 \\
SVC & 0.493 ± 0.144 & 1.875 ± 0.379 & 598.998 ± 50.066 & 25.025 ± 0.400 \\
\hline
\end{tabular}
\caption{Average execution times (in seconds) for the different machine learning models using the colored noises time series data for microstates with size $N=2$, $N=3$ and $N=4$, as well as for the raw time series data. 
The results consider all 20 possible training and testing permutations across 5 distinct datasets. 
The values represent the mean and standard deviation of execution time for each model.}
\label{tab:colored_noises}
\end{table*}

\end{appendices}

\end{document}